# Aerodynamic Prediction of a CRM High-lift Configuration using a modified three equation turbulence model

Shaoguang Zhang, [1,*] Haoran Li [2,†], Yufei Zhang [1,‡]

(1. Tsinghua University, Beijing, 100084, People's Republic of China

(2. Beijing Institute of Mechanical and Electrical Engineering, Beijing, 100084, People's Republic of China)

**Aerodynamic simulations were carried out in the study presented in this paper focusing on the stall performance of the High-Lift Common Research Model obtained from the fourth AIAA High-Lift Prediction Workshop. Various turbulence models of Reynolds-average Navier–Stokes simulations are analyzed. A modified version of the transitional $k - \overline{v^2} - \omega$ model developed to enhance stall prediction accuracy for high-lift configurations with a nacelle chine. The vortex generator, three-element airfoil, and high-lift model are numerically simulated. The results reveal that implementing a $k - \overline{v^2} - \omega$ model with separation shear layer fixed notably enhances the stall prediction behavior for both the three-element airfoil and high-lift configuration without affecting the prediction of the vortex strength of a vortex generator. Moreover, incorporating rotation correction into the SPF $k - \overline{v^2} - \omega$ model improves the prediction of vortex strength and further enhances stall prediction for the high-lift configuration. The relative error in predicting the maximum lift coefficient is less than 5% of the experimental data. The study also investigated the impact of the nacelle chine on the stall behavior of the high-lift configuration. The results demonstrate that the chine vortex can mitigate the adverse effects of the nacelle/pylon vortex system and increase the maximum lift coefficient.**

---

[*] Ph. D student, School of Aerospace Engineering, email: zsg21@mails.tsinghua.edu.cn
[†] Engineer, Beijing Institute of Mechanical and Electrical Engineering, email: lihr17@tsinghua.org.cn
[‡] Associate professor, School of Aerospace Engineering, senior member AIAA, email: zhangyufei@tsinghua.edu.cn (corresponding author)

## Nomenclature

(Nomenclature entries should have the units identified)

| | | |
|---|---|---|
| $C_L$ | = | lift coefficient |
| $C_{L,max}$ | = | maximum lift coefficient |
| $Cm$ | = | pitching moment coefficient |
| $C_p$ | = | pressure coefficient |
| $\rho$ | = | air density, kg/m³ |
| $Re$ | = | Reynolds number |
| $S$ | = | shear rate, 1/s |
| $\mu$ | = | molecular viscosity, Pa.s |
| $\nu$ | = | kinematic molecular viscosity, $\mu/\rho$, $m^2/s$ |
| $\Omega$ | = | magnitude of vorticity, $1/s$ |
| $C_{p0}$ | = | stagnation pressure coefficient |
| $\delta$ | = | boundary-layer thickness, mm |
| $e$ | = | vortex generator chord length, mm |
| TI | = | turbulence intensity |

## I. Introduction

The design of high-lift configurations significantly impacts the safety and efficiency of commercial aircraft. Engineers primarily aim to enhance the maximum lift coefficient when working on high-lift configurations. Meredith indicated that a mere 1% increase in the maximum lift coefficient can turn into a payload increase of 22 passengers for a large civil aircraft with a fixed approach speed during landing [1]. Various methods have been proposed to improve the maximum lift coefficient of high-lift configurations, such as implementing double-slotted or triple-slotted flaps and increasing slat deflection[2]. Nacelle chines [[3], 4, 5] offer a straightforward passive flow control approach for enhancing high-lift performance. However, these enhancements often introduce additional complexity to the high-lift configuration structure. The intricate geometry of these configurations leads to complex flow patterns, including separated flows at high angle of attack values, transitions between laminar and turbulent flow on the leading edges of wing elements, wake and boundary-layer interactions, and the formation of wing-tip vortices [6]. Current computational fluid dynamics (CFD) tools need to accurately predict transitions and separated flows to model such intricate flow phenomena. Accurately predicting the aerodynamic performance of high-lift configurations remains a challenging engineering problem, especially the stall behavior [7].

The American Institute of Aeronautics and Astronautics (AIAA) has organized four high-lift prediction workshops. Each workshop provided a diverse high-lift model and reliable experimental data for CFD validation. Researchers have identified the turbulence model as a pivotal element in predicting the stall behavior of high-lift configurations through a statistical analysis of the workshop outcomes. Most turbulence models used in predicting high-lift flows rely on a linear constitutive model of the Reynolds stress tensor [8]. Among the turbulence models utilized by workshop participants, the Spalart-Allmaras (SA) [9] and shear stress transport (SST) [10] models are commonly employed in high-lift configuration design. However, it has been reported that these models struggle to accurately predict the stall behavior of high-lift configurations [[11], 12]. Various modifications have been introduced to enhance their ability to predict high-lift aerodynamics. For example, Hellsten [8] developed a new $k - \omega$ model tailored to high-lift flow requirements using an explicit algebraic Reynolds-stress model as the constitutive relation between the Reynolds stress tensor and mean-velocity gradient. This new model exhibited improved performance in predicting high-lift airfoil aerodynamics. Langlois et al. [13] found that adding curvature correction to the turbulence model improved the accuracy of separation region prediction for the JAXA Standard Model (JSM). Yasushi et al. [14] investigated how the quadratic constitutive relation (QCR) model proposed by Spalart [15] affected high-lift flow predictions. They observed that the SA model with QCR underpredicted the maximum lift coefficient of the JSM and led to premature flow separation on the outboard of the wing.

Flow separation is a common occurrence on wing surfaces prior to stall. When turbulence models fail to predict separated flow accurately, it can result in an inaccurate stall prediction. Rumsey [16] indicates that the inaccuracies in turbulence models stem from an underestimation of shear stress in the shear layer. A "separation fix" was introduced to the $k - \omega$ model to address this issue using the production-to-dissipation ratio ($P_k/\varepsilon$) of turbulent kinetic energy as a trigger [16]. This modified turbulence model yielded accurate predictions for scenarios such as 2-D hills and 2-D hump flows. Li et al. [[17], 18] developed a new separating shear-layer fixed (SPF) $k - \overline{v^2} - \omega$ model to simulate the stall behavior of iced airfoils. Their numerical tests showed a maximum relative error of approximately 5% when compared to experimental values on various iced airfoils. The original $k - \overline{v^2} - \omega$ model proposed by Lopez and Walters in 2016[19] performs well in various boundary-layer transition and free-shear-flow cases. Zhang et al. [20] conducted research to validate different turbulence models in predicting stall behavior in high-lift configurations. They concluded that accurately modeling the free shear layers might be a pivotal factor in precisely predicting high-lift stall behavior.

In model corrections for predicting separated flow, a common approach involves influencing eddy viscosity by either increasing the $\omega$ destruction term or enhancing the $k$ production term. This method was also employed in the unpublished Generalized $k - \omega$ (GEKO) turbulence model introduced by Menter in 2019 [21]. The GEKO model incorporates three functions ($F_1, F_2, F_3$) to modify the production, destruction, and cross-diffusion terms in the $\omega$ equation to change the eddy viscosity in the boundary layer and free shear layer, which aims to enhance the prediction of separated flow [21]. Furthermore, improving the curvature correction fix involves increasing the destruction term in the $\omega$ equation to change eddy viscosity. This correction method is widely used in data-driven turbulence modeling approaches. Recently, Yan et al. [22] introduced a spatially distributed factor $\beta$ to control eddy viscosity production in the SA model. Their results highlighted that the primary error in the SA model's prediction of separated flow originates from the separated shear layer. This conclusion is consistent with the results of the SPF $k - \overline{v^2} - \omega$ model [17]. Recently, Wu et al. [23] employed a combination of field inversion and machine learning along with symbolic regression to provide an analytical relationship between the correction factor $\beta$ of the baseline turbulence model and local flow variables. The newly developed SST-SR model has demonstrated excellent predictive performance in cases such as the Ahmed body and periodic hill simulations [23].

In this study, the $k - \overline{v^2} - \omega$ model with separation-layer correction and rotation correction is employed to predict the stall behavior of the CRM-HL configuration. Initially, the impact of these two corrections on the model's ability is assessed to predict the vortex strength of a vortex generator. Then, the influence of the shear layer fixed and rotation corrections on the stall behavior prediction of the multielement airfoil and high-lift configuration is investigated. Finally, the flow mechanism of the nacelle chine on improving the stall behavior of the high-lift configuration is studied.

## II. Numerical Method

### A. Numerical solver

CFL3D version 6.7 [24]is used to predict the stall behavior of the CRM-HL configuration in this study. The flux-difference-splitting technique of Roe[25] is adopted for spatial discretization. The MUSCL (Monotone Upstream-centered Scheme for Conservation Laws) approach of van Leer[26] is used to determine state-variable interpolations at the cell interfaces. The three equations of the turbulence model are solved decoupled using the implicit approximate factorization method. The advective terms in turbulence equations are discretized using first-order upwinding scheme. The production terms of $k$, $\overline{v^2}$, and $\omega$ are treated explicitly (lagging in time). The destruction terms and additional cross-diffusion term in $\omega$ equation are treated implicitly, which may help increase the diagonal dominance of the left-hand-side matrix. The implicit approximate-factorization method is used for time advancement. Multigrid and mesh sequencing are provided to accelerate the convergence. This solver supports multiple-zone grids connected in a one-to-one, patched, or overset manner. In study presented in this paper, the one-to-one manner grid is used.

**B. Turbulence Model**

In this section, a modified version of the $k - \overline{v^2} - \omega$ model is introduced incorporating a separated shear layer fix (SPF) to increase the eddy viscosity in separated regions. The original $k - \overline{v^2} - \omega$ model was initially proposed by Lopez and Walters in 2016 [19] and consists of three transport equations: total fluctuation energy ($k$), fully turbulent fluctuation energy ($\overline{v^2}$), and specific dissipation rate ($\omega$). These transport equations are presented in equations (1-3). The $R_{NAT}$ and $R_{BP}$ terms in this model are utilized to represent the natural and bypass transition processes, respectively. For detailed information on each term of this model, readers can refer to reference [17].

$$\frac{\partial k}{\partial t} + u_j \frac{\partial k}{\partial x_j} = \frac{1}{\rho} P_k - min(\omega k, \omega \overline{v^2}) - \frac{1}{\rho} D_k + \frac{1}{\rho}\frac{\partial}{\partial x_j}\left[\left(\mu + \frac{\rho \alpha_T}{\sigma_k}\right)\frac{\partial k}{\partial x_j}\right] \quad (1)$$

$$\frac{\partial \overline{v^2}}{\partial t} + u_j \frac{\partial \overline{v^2}}{\partial x_j} = \frac{1}{\rho} P_{\overline{v^2}} + R_{BP} + R_{NAT} - \omega \overline{v^2} - \frac{1}{\rho} D_{\overline{v^2}} + \frac{1}{\rho}\frac{\partial}{\partial x_j}\left[\left(\mu + \frac{\rho \alpha_T}{\sigma_k}\right)\frac{\partial \overline{v^2}}{\partial x_j}\right] \quad (2)$$

$$\frac{\partial \omega}{\partial t} + u_j \frac{\partial \omega}{\partial x_j} = \frac{1}{\rho} P_\omega + \left(\frac{C_{\omega R}}{f_W} - 1\right)\frac{\omega}{\overline{v^2}}(R_{BP} + R_{NAT}) - f_{RC}f_{NE}C_{\omega 2}\omega^2 f_W^2$$
$$+ 2\beta^*(1 - F_1^*)\sigma_{\omega 2}\frac{1}{\omega}\frac{\partial k}{\partial x_j}\frac{\partial \omega}{\partial x_j} + \frac{1}{\rho}\frac{\partial}{\partial x_j}\left[\left(\mu + \frac{\rho \alpha_T}{\sigma_\omega}\right)\frac{\partial \omega}{\partial x_j}\right] \quad (3)$$

The SPF modification is achieved by utilizing a coefficient $f_{NE}$ in Eq. (3) on the destruction term of the $\omega$ transport equation. The production-to-dissipation ratio $P_{\overline{v^2}}/\varepsilon$ of the turbulent kinetic energy is used as a trigger of this modification. The shear layer modification region is identified by the switch function $\Gamma_{SSL}$, where $P_{\overline{v^2}}/\varepsilon > 2.5$. The modification term $f_{NE}$ is turned off where $P_{\overline{v^2}}/\varepsilon$ is less than 2.5, and the model reverts to the original $k - \overline{v^2} - \omega$ model [17, 18]. The $Re_\Omega$ term in Eq. (4) is used to determine the magnitude of the modification, which indicates that the modification is enlarged in the large vorticity region away from the wall. The maximum value of $f_{NE}$ is chosen as 3.3 to remain bounded.

$$f_{NE} = min(max(300Re_\Omega \Gamma_{SSL}, 1), 3.3), \quad Re_\Omega = \frac{d^2\Omega}{\nu}, \quad C_{\omega 2} = 0.92 \tag{4}$$

$$\Gamma_{SSL} = \frac{1}{1 + e^{-10(\frac{P_{\overline{v^2}}}{\varepsilon} - C_{SSL})}}, \quad \frac{P_{\overline{v^2}}}{\varepsilon} = \frac{\mu_{T,s} S^2}{\rho \overline{v^2} \omega}, \quad C_{SSL} = 2.5 \tag{5}$$

The production terms are expressed as:

$$P_k = \nu_T S^2, \quad P_{\overline{v^2}} = \nu_{T,s} S^2, \quad P_\omega = (C_{\omega 1} \frac{\omega}{\overline{v^2}} \nu_{T,s}) S^2, \quad C_{\omega 1} = 0.44 \tag{6}$$

The SPF model has been tested on iced airfoils, multiple-element airfoils and complex three-dimensional high-lift configurations [[17], 18, 20]. Zhang et al.[20] indicated that the $k - \overline{v^2} - \omega$ model coefficient $C_{\omega 2}/C_{\omega 1}$ is related to the variation tendency of the production-to-dissipation ratio $P_{\overline{v^2}}/\varepsilon$ and influences the transport behavior of the turbulence kinetic energy. The model coefficient $C_{\omega 2}/C_{\omega 1}$ has a strong influence on accurately predicting the stall behavior of the high-lift configuration.

The $k$ equation lacks a Coriolis force term, making it incapable of considering the impact of rotation [27]. As a result, ad hoc coordinate-frame-rotation adjustments are necessary for the eddy viscosity model. The SA model with rotation and curvature corrections was initially introduced by Spalart and Shur in 1997 and is referred to as SARC [[28], 29]. Smirnov and Menter extended the rotation correction to a two-equation model and developed the SST model with a rotation-corrected version, known as SST-RC[30]. Both the SARC and SST-RC models demonstrate enhanced performance in predicting vortex wake of vortex generators[28]. Hellsten redefined the Richardson number and proposed a simplified form of rotation correction [31]. This simplified form of the rotation correction is employed in this study. The destruction term is modified by the rotation correction; see $f_{RC}$ in Eq. (3) as:

$$f_{RC} = \frac{1}{1 + C_{RC} Ri} \tag{7}$$

The Richardson number in Eq. (8) can be written as:

$$Ri = \frac{W}{S}(\frac{W}{S} - 1) \tag{8}$$

$$S = \sqrt{2S_{ij}S_{ij}} \qquad S_{ij} = \frac{1}{2}(\frac{\partial u_i}{\partial x_j} + \frac{\partial u_j}{\partial x_i})$$

$$W = \sqrt{2W_{ij}W_{ij}} \qquad W_{ij} = \frac{1}{2}(\frac{\partial u_i}{\partial x_j} - \frac{\partial u_j}{\partial x_i})$$

The value of the constant $C_{RC}$ is taken as 1.4.

## III. Test Cases

### A. Vortex generator

This section aims to test the effect of the separated shear layer correction on vortex intensity and examine factors affecting the prediction accuracy of a high-lift configuration. A conventional vortex generator is chosen as the first test case[32]. The model has been experimentally tested in the Langley 20- by 28-Inch Shear Tunnel. The geometry of the vortex generator is illustrated in Fig. 1 The boundary-layer thickness ($\delta$) is 35 mm at the location of the vortex generator. The freestream velocity is 34 m/s. The height of the vortex generator is 35 mm, and its length ($e$) is 70 mm.

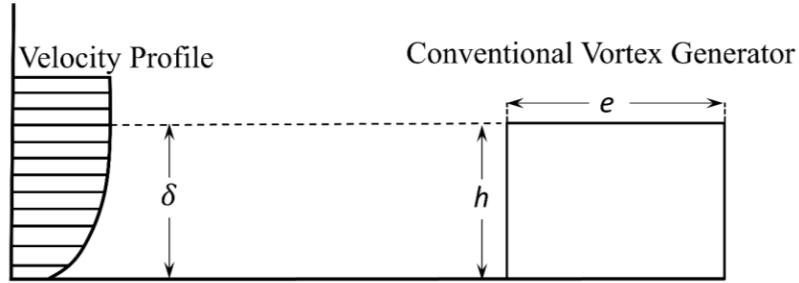

Fig. 1 Geometry of the conventional vortex generator

The computational domain and the grid used in this work are presented in Fig. 2. The leading edge of the vortex generator is located at $x = 0$. The inflow boundary is at $x = -128h$, which can ensure that the boundary layer thickness near the leading edge of the vortex generator is the same as the height of the vortex generator. The outflow boundary is at $x = 516h$, which provides sufficient distance to test the vortex wake.

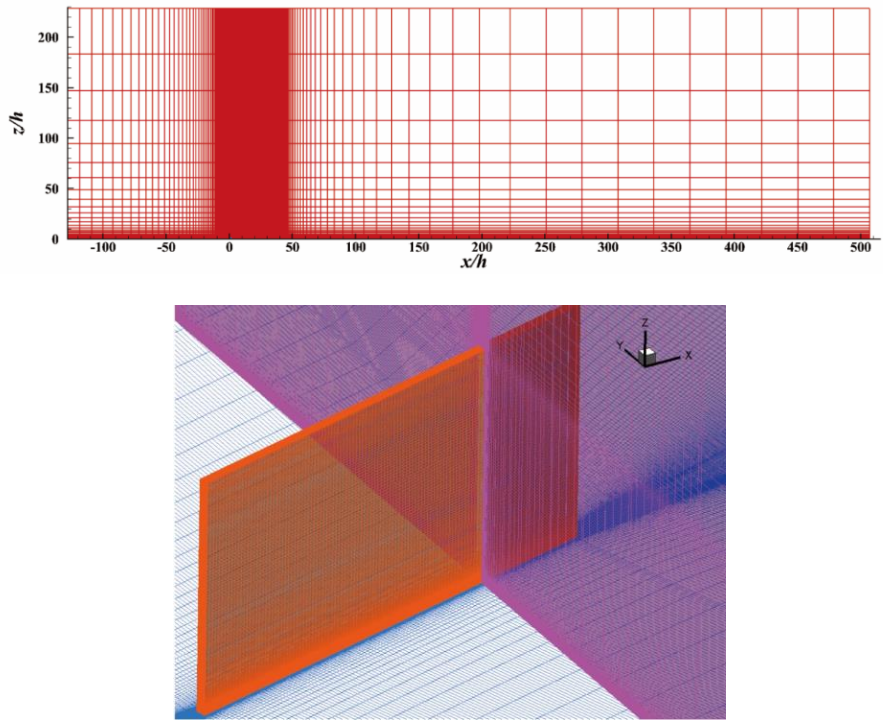

**Fig. 2 Computational domain and grid of the conventional vortex generator**

A grid convergence study is conducted with three grid sets, which are named coarse, medium, and fine. The grid cell numbers are 14 million, 24 million, and 42 million, respectively. The medium grid set is obtained by increasing the grid points in each direction of the coarse grid set by 1.3 times. The fine grid set is similarly generated based on the medium grid set. The first-layer $\Delta y^+$ values for the three grid sets are approximately 1.0, 0.6, and 0.4, respectively. The grid spacing growth rates of the boundary layers are 1.15 and 1.2 for the other regions, respectively.

The peak value of the streamwise vorticity ($\omega_{x,max}$) is chosen to quantify the evolution of the vortex wake. Fig. 3 compares the predicted $\omega_{x,max}$ at AOA = 16 deg using the SPF $k - \overline{v^2} - \omega$ model with the experimental data. The $\omega_{x,max}$ values predicted by the medium and fine grids are virtually identical, which indicates grid convergence in the vortex generator case. However, the peak values of streamwise vorticity at different streamwise locations remain underpredicted. Subsequent calculations are performed using the fine grid to test the accuracy of different turbulence models.

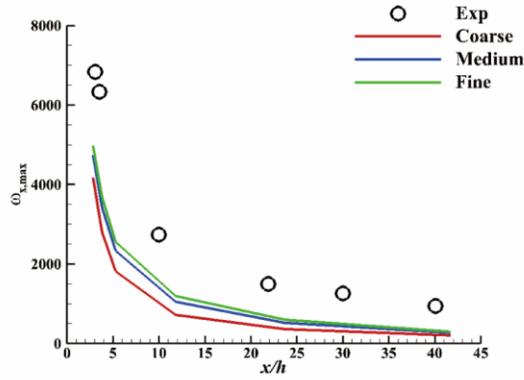

**Fig. 3 Comparisons of $\omega_{x,max}$ predicted by the SPF $k - \overline{v^2} - \omega$ model using three sets of grids**

Figure 4 presents a comparison of the $\omega_{x,max}$ values predicted by different turbulence models. When comparing the original $k - \overline{v^2} - \omega$ and SPF $k - \overline{v^2} - \omega$ models, it is evident that the separation shear layer correction does not affect the model's ability to predict the peak streamwise vorticity. In contrast, both the $k - \overline{v^2} - \omega - RC$ and SPF-RC models with rotation correction predict peak streamwise vorticity values notably consistent with the experimental data. This observation highlights the enhancement brought by rotation correction in predicting the streamwise vortex wake. However, the peak values of streamwise vorticity downstream are still slightly underpredicted.

Streamwise vorticity contours at 3 downstream stations are presented in Fig. 5. The experimental results show that the vortex core gradually moves in the positive y-axis direction when the vortex develops downstream. The CFD method captures the same tendency. The rotation correction improves the prediction accuracy of the streamwise vortex strength but still slightly underestimates the regions with higher streamwise vortex intensity far downstream.

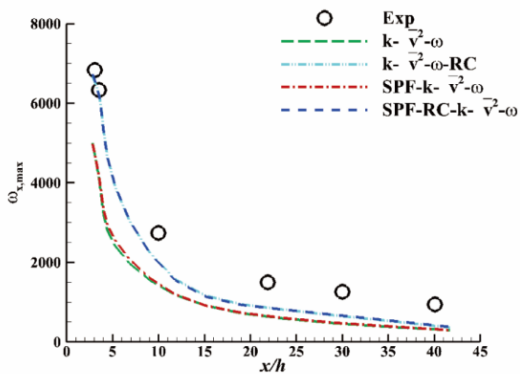

**Fig. 4 Comparisons of $\omega_{x,max}$ predicted by different turbulence models**

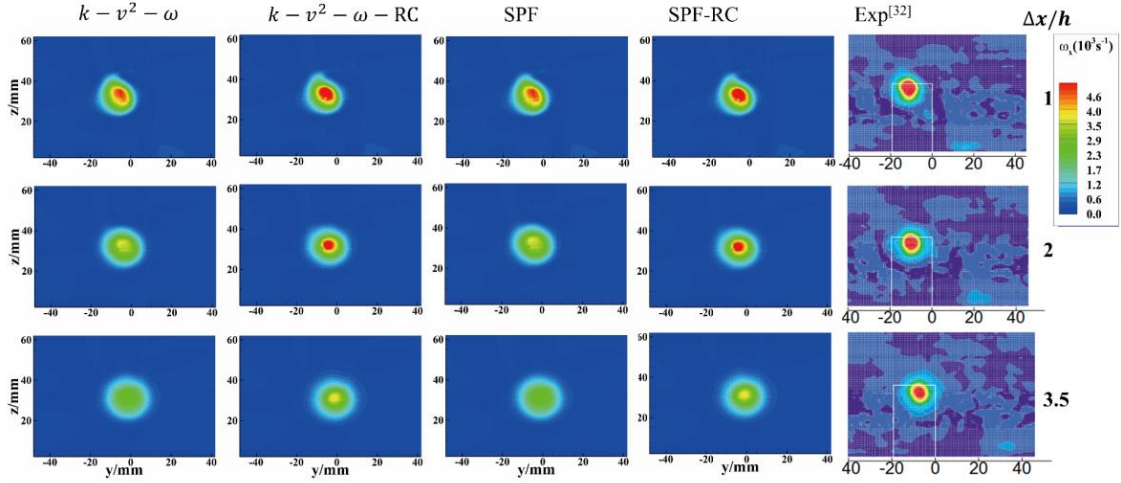

**Fig. 5 Streamwise vorticity contours at 3 downstream stations of the vortex generator for $AOA = 16°$**

**B. 30P30N Multielement Airfoil**

The 30P30N multielement airfoil is chosen as the second test case to validate the effects of separated shear layer correction and rotational correction on predicting the stall performance. The airfoil was designed by McDonnell Douglas. Detailed experimental data were provided by NASA Langley Research Center's Low Turbulence Pressure Tunnel (LTPT) [33], including aerodynamic coefficients and pressure distribution. The grid in this study is generated based on the fine grid described in reference [39]. Refinement on the upper surface of the airfoil is adopted, as shown in Fig. 6. The first grid layer has a height of $5.0 \times 10^{-6}$ to ensure that $\Delta y^+$ is less than 1.0.

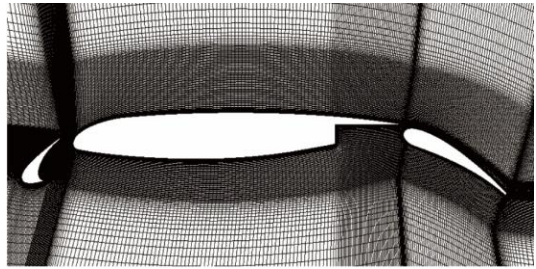

**Fig. 6 Computational grid of the 30P30N multielement airfoil**

The accurate prediction of transition by the $k - \overline{v^2} - \omega$ and SPF models and its impact on predicting stall in the 30P30N configuration is discussed here. The transition process can be modeled through the $\overline{v^2}$ transport equation in the $k - \overline{v^2} - \omega$ model. The $R_{NAT}$ and $R_{NAT}$ terms in the $\overline{v^2}$ equation represent natural and bypass transition processes.

10/40

Wind tunnel experiments for the 30P30N configuration were conducted at the low-turbulence pressure tunnel of NASA Langley. The value of free stream turbulence intensity is 0.03%. The SA and SST models are used for fully turbulent calculations. Friction coefficient curves obtained from different turbulence models are shown in Fig. 7. The experimentally measured transition point on the main wing is around x/c=0.05 [34]. The original $k - \overline{v^2} - \omega$ and SPF models accurately predict the transition location on the main wing when the turbulence intensity is 0.03%. These two models accurately predict flap transition locations but underestimate the peak friction coefficient. When turbulence intensity reaches about 0.1%, the transition points on the main wing and flaps shift towards the front. Increasing turbulence intensity from 0.03% to 0.1% does not greatly affect the $k - \overline{v^2} - \omega$ and SPF models' predictions of the maximum lift coefficient and stall behavior. However, it can avoid the computational instability caused by laminar separation bubbles. Therefore, a turbulence intensity of 0.1% was chosen for subsequent calculations.

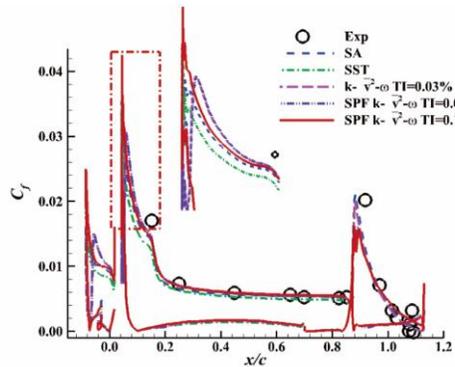

Fig. 7  Skin friction distribution at $AOA = 8°$ and a Reynolds number of 5 million

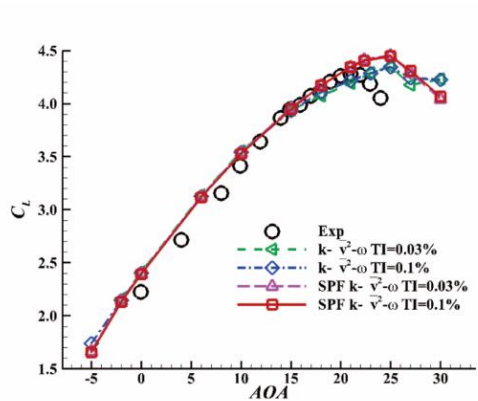

Fig. 8  Comparison of lift coefficient curves predicted by the $k - \overline{v^2} - \omega$ model and the SPF model under different turbulence intensities，$Re = 5 \times 10^6$

Aerodynamic coefficients predicted by different turbulence models of the 30P30N configuration are shown in Fig. 9. The SST and original $k - \overline{v^2} - \omega$ models underpredict the maximum lift coefficient, mainly because they underestimate the loads on the main wing and flap, as shown in Fig. 9(b) and Fig. 9(c). The lift curve predicted by the SA model agrees well with the experimental data. However, the SA model overpredicts the maximum lift coefficient with the same solver in the study by Rumsey et al [34]. The factors contributing to different results may be due to differences in grid topology or solution strategies. The SPF model significantly improves the prediction of the maximum lift coefficient and the loads on the main wing and flap compared with the original $k - \overline{v^2} - \omega$ model. This suggests that the separated shear layer correction in the model proves advantageous in forecasting the stall characteristics of multiple-element airfoils, which is consistent with our prior research[20]. It also highlights the benefits of separated shear layer correction in predicting stall within high-lift configurations with flow separation. Upon comparing results with and without rotational correction, it is observed that rotational correction does not impact the model's ability to predict stall characteristics for this two-dimensional geometry. The SA, SPF $k - \overline{v^2} - \omega$ and SPF-RC $k - \overline{v^2} - \omega$ models yield favorable results in terms of pressure distribution at AOA=21 degrees, as shown in Fig. 9(b). However, the SST and original $k - \overline{v^2} - \omega$ models fall short in predicting the suction peak on the main element.

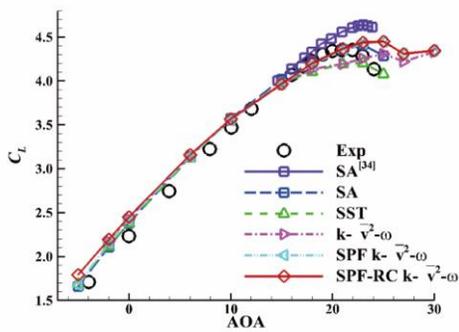
a) $C_{L,total}$ vs AOA

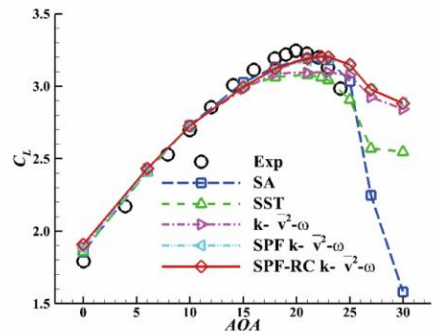
b) $C_{L,mainwing}$ vs AOA

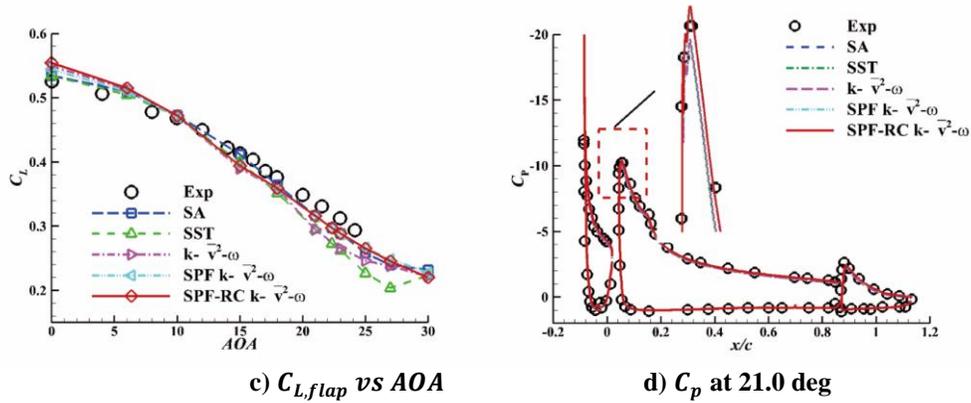

c) $C_{L,flap}$ vs AOA  
d) $C_p$ at 21.0 deg

**Fig. 9 Aerodynamic coefficients predicted by different turbulence models of the 30P30N three-element airfoil**

Fig. 8 shows the nondimensional streamwise velocity $U/U_{inf}$ contours predicted by different models at $AOA = 21°$. Two distinct low-speed regions are observed above the flap: the slat wake and the main element wake. These regions give rise to shear layers both above and below due to substantial velocity gradients. The most pronounced differences in velocity contours predicted by the different turbulence models are concentrated within the wake region. Flow-reversal above the flap is predicted by the SST and the original $k - \overline{v^2} - \omega$ models. These two models often overpredict the width of wakes and underestimate the velocity within the wake area, as shown in Fig. 11. The locations for surface-normal profile measurements in Fig. 11 are detailed in reference [36]. The SA model accurately predicts the main wing wake but underestimates the velocity of the slat wake. The SPF $k - \overline{v^2} - \omega$ model predicts a smaller and closer main wing wake to the surface of the flap compared to the original model. The rotation correction in the SPF $k - \overline{v^2} - \omega$ model has a negligible effect on the predicted velocity field.

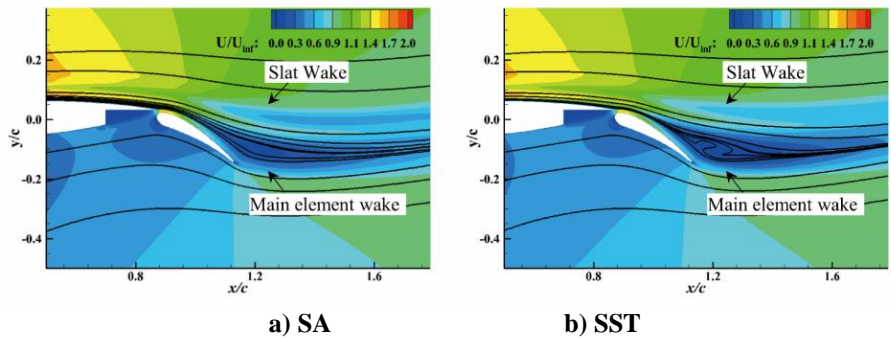

a) SA  
b) SST

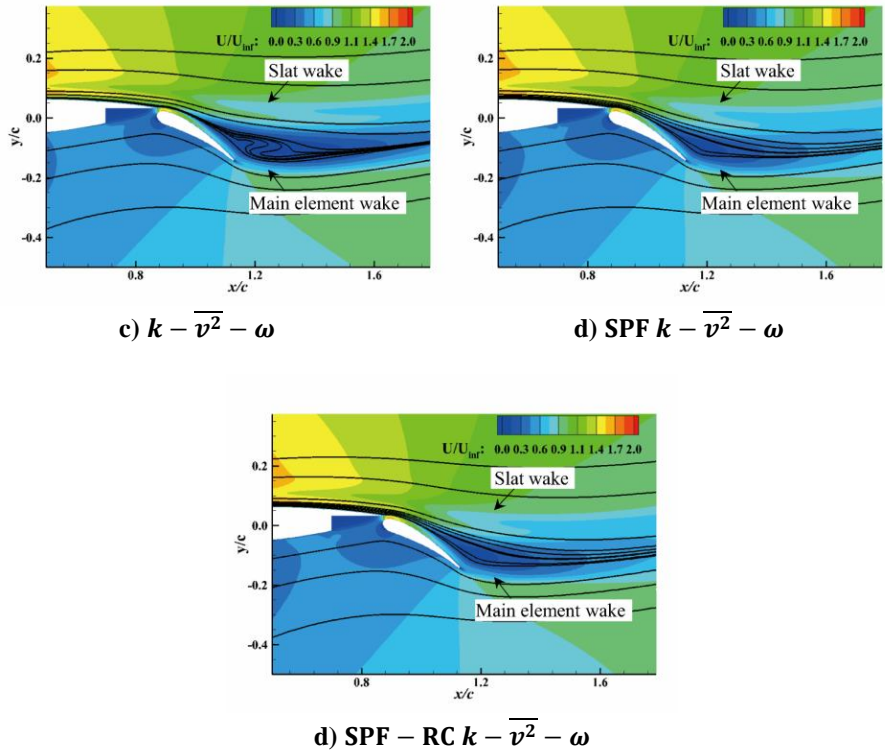

c) $k - \overline{v^2} - \omega$  d) SPF $k - \overline{v^2} - \omega$

d) SPF $-$ RC $k - \overline{v^2} - \omega$

Fig. 10 Nondimensional streamwise velocity $U/U_{inf}$ contours predicted by different turbulence models, AOA=21.0 deg

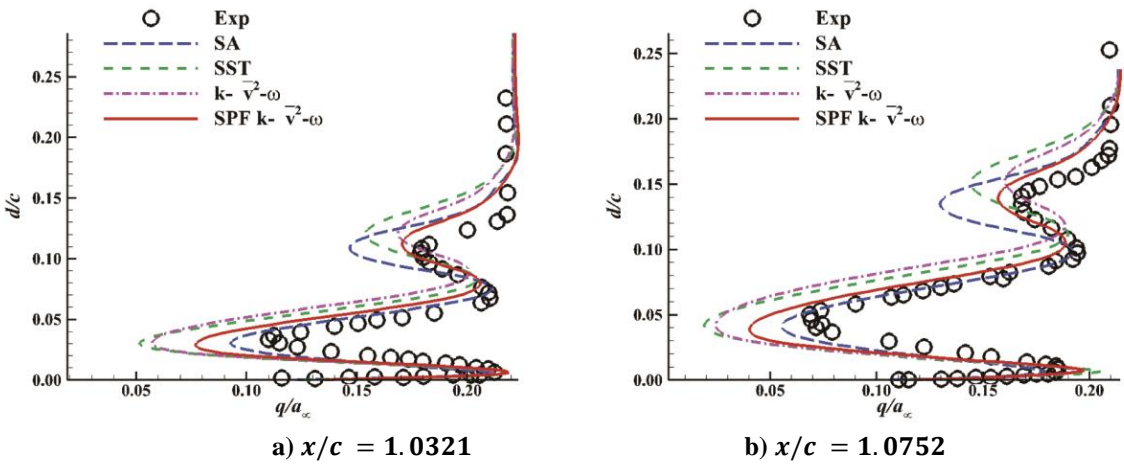

a) $x/c = 1.0321$  b) $x/c = 1.0752$

Fig. 11 Velocity profile predicted by different turbulence model for $AOA = 21°$

The 30P30N configuration tends to exhibit off-body recirculation at high angle of attack. The expansion of the separation leads to stall [37]. The streamwise velocity contour at $AOA = 22.34°$ is shown in Fig. 9. The SST and original $k - \overline{v^2} - \omega$ predict flow-reversal above the flap, but underpredict the maximum lift coefficient. Flow-reversal above the flap is not predicted by the SA, SPF and the SPF-RC models. However, they provide more accurate predictions of aerodynamic forces. This phenomenon indicates that the currently improved models still struggle to accurately represent the flow structure post-stall.

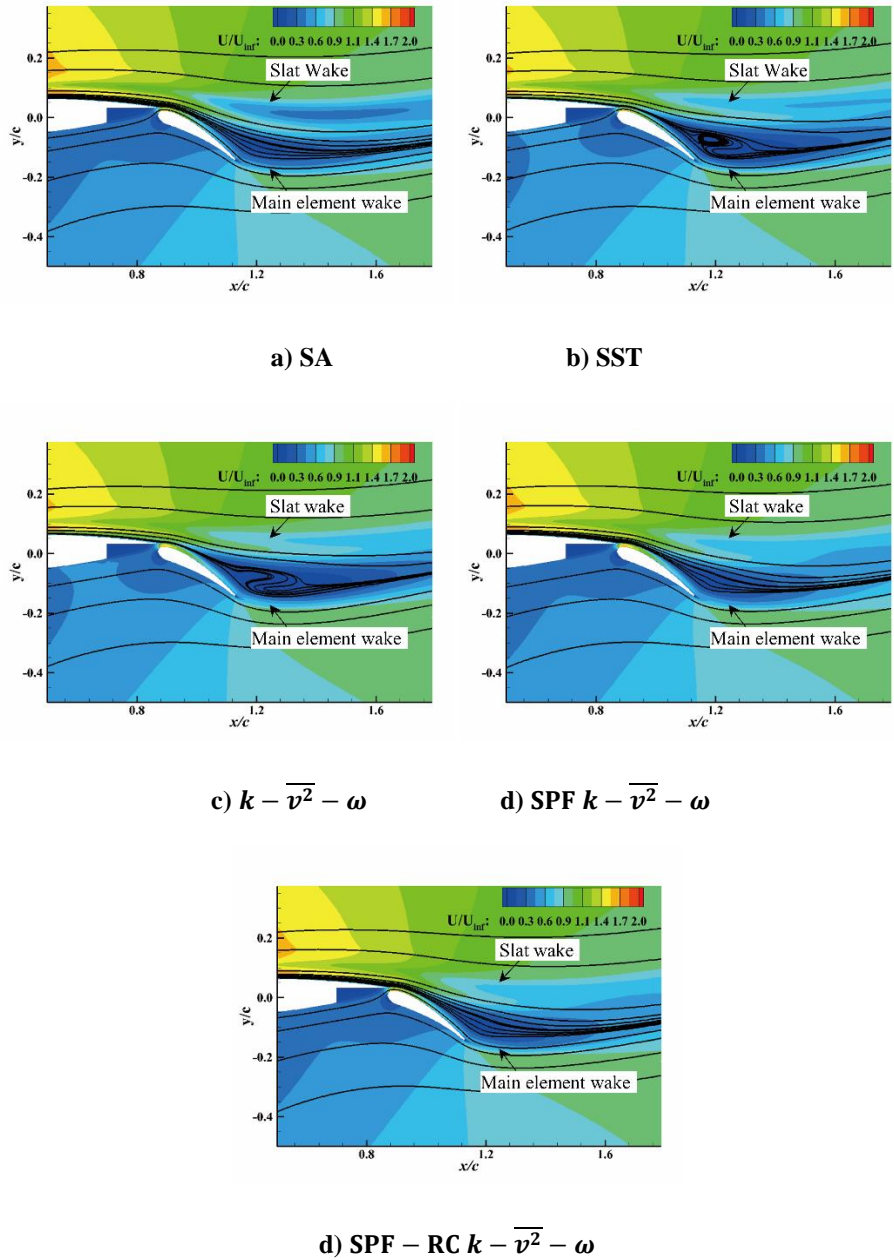

a) SA

b) SST

c) $k - \overline{v^2} - \omega$

d) SPF $k - \overline{v^2} - \omega$

d) SPF $-$ RC $k - \overline{v^2} - \omega$

**Fig. 12 Nondimensional streamwise velocity $U/U_{inf}$ contours predicted by different turbulence models, AOA=22.34 deg**

Fig. 13 presents comparisons of the $P_{\overline{v^2}}/\varepsilon$ contours predicted by different turbulence models. The ratio of turbulent kinetic energy generation to dissipation can be obtained using Eq. (5). A significant increase in $P_{\overline{v^2}}/\varepsilon$ values are seen at the shear layer when predicted by the $k - \overline{v^2} - \omega$ model. The value of $P_{\overline{v^2}}/\varepsilon$ exceeds the threshold of 2.5 in the free shear layer at some locations; consequently, the modification for the separated shear layer will be turned on according to Eq. (5). Fig. 13(b) illustrates the notable increase in $P_{\overline{v^2}}/\varepsilon$ values predicted by the SPF model for the shear layer. Adding rotation correction to the SPF $k - \overline{v^2} - \omega$ model results in a slight increase in $P_{\overline{v^2}}/\varepsilon$ values at the shear layer, which is aligned with the expectations from Eq. (8). The maximum value of $f_{RC}$ is approximately 1.54 within the shear layer where $S \gg W$, resulting in a relatively limited impact on the $P_{\overline{v^2}}/\varepsilon$ increase. This observation elucidates why the rotation correction has minimal effect on the SPF model's prediction of multielement airfoil stall in the 30P30N configuration.

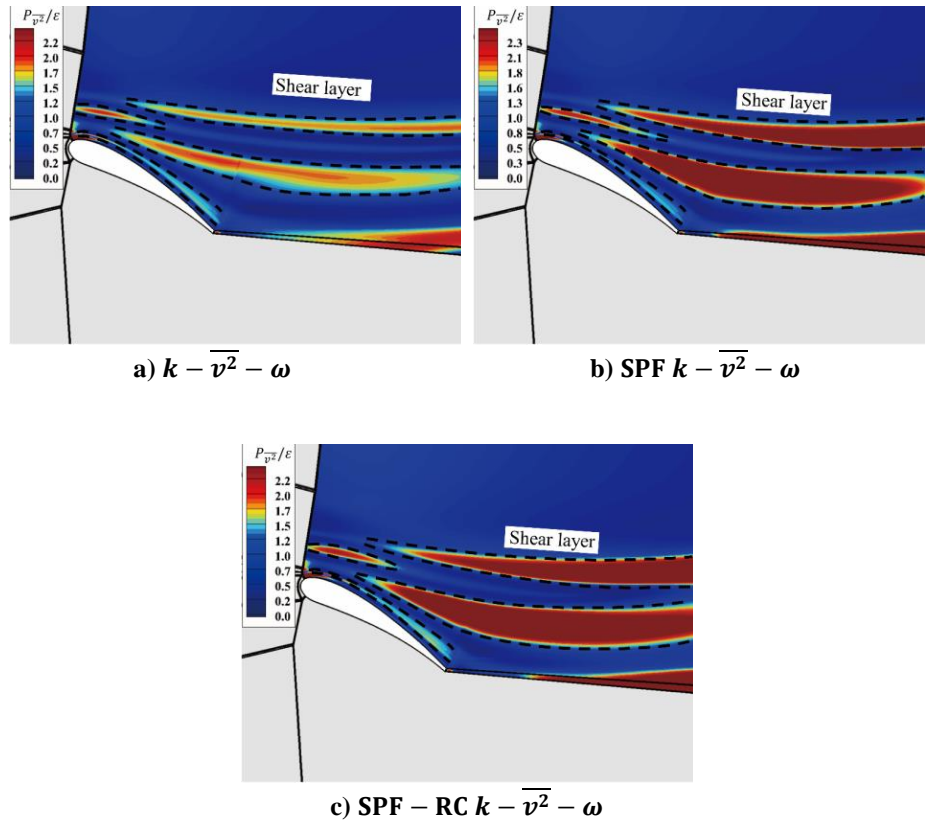

a) $k - \overline{v^2} - \omega$     b) SPF $k - \overline{v^2} - \omega$

c) SPF − RC $k - \overline{v^2} - \omega$

Fig. 13 Comparisons of the $P_{\overline{v^2}}/\varepsilon$ contour predicted by different turbulence models

## C. The High-lift Version of the Common Research Model

*1. Test case description*

The high-lift version of the Common Research Model (CRM-HL) was used as the standard model in the fourth High-Lift Prediction Workshop [40]. The CRM-HL wind tunnel model closely replicates the components of a modern transport aircraft, featuring a fuselage, a main wing, leading edge slats, a flow-through nacelle, a nacelle pylon, trailing edge flaps, and flap track fairings, as depicted in Fig. 10. Rigorous wind tunnel tests were conducted on the CRM-HL configuration at the QimetiQ5m wind tunnel in Farnborough [44]. The organizers furnished us with high-quality test results, including aerodynamic forces, moments, pressure distributions at various span locations, and oil flow visualization. Three configurations with different flap deflection angles were provided by the workshop. The nominal configuration [40] is chosen as the test case in this paper, for which the inboard and outboard flap deflection angles are 40 degrees and 37 degrees, respectively. The key dimensions and parameters are summarized in Table 1. The Reynolds number based on the mean aerodynamic chord is 5.49 million, and the freestream Mach number is 0.2. The mean longitudinal turbulence intensity of the incoming flow stands at 0.08%.

**Table 1 Main dimensions of the nominal CRM-HL configuration**

| Main dimension | | |
|---|---|---|
| Half span, s | [m] | 2.938 |
| Wing reference area, A/2 | [m$^2$] | 1.918 |
| Reference chord, c$_{ref}$ | [m] | 0.7 |
| Aspect ratio, $\Lambda$ | [-] | 9 |
| Slat deflection angle, $\delta_s$ | [°] | 30 |
| Inboard Flap deflection angle, $\delta_f$ | [°] | 40 |
| Outboard Flap deflection angle, $\delta_f$ | [°] | 37 |

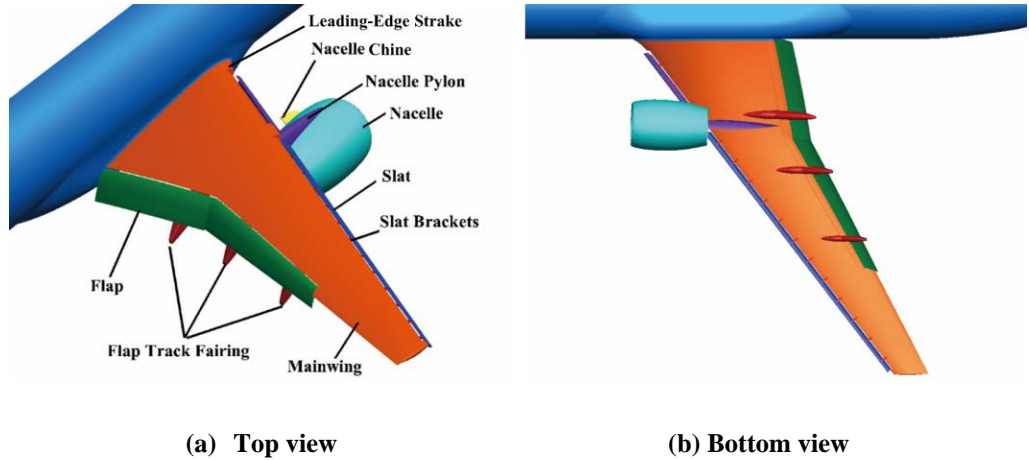

(a) Top view            (b) Bottom view

Fig. 14 CRM-HL Wind tunnel model

*2. Numerical settings and computational strategy*

This section outlines the setup and strategies for numerical calculations. It uses a steady-state method to solve the flow field at all incidence angles without sub-iterations. The value of free stream turbulence intensity is 0.1%. The process starts with a uniform initial field. The initial value of CFL number is selected as 0.05, and it ramped over 1000 steps to CFL = 1.0. Multigrid acceleration is used to accelerate the convergence, typically involving 3000 computation steps on the coarsest grid level, 2000 on the next coarser level, and at least 5000 on the baseline until convergence is reached.

*3. Computational grid and grid convergence study*

The CRM-HL has more complex components than the JAXA Standard Model (JSM) provided by the third High-lift Prediction Workshop [40] A one-to-one structural grid of the JSM model was generated in our previous work [20]. In this work, a structural one-to-one multiblock grid of the CRM-HL configuration is also adopted. Compared with the JSM configuration, the CRM-HL configuration has a nacelle chine and more slat brackets, which makes it more difficult to generate a structural grid.

A grid convergence study is investigated on the lift coefficient, drag coefficient, and pitching moment coefficient using the SPF $k - \overline{v^2} - \omega$ model. Our analysis includes five sets of grids, labeled coarse, medium, medium fine, fine and extra fine, as shown in Fig. 15. The detailed information of different grids is shown in Table 2. The coarse grid consists of a total of 78 million cells. The medium grid is created by increasing the grid number in each direction of the blocks by approximately 1.25 times in relation to the coarse grid. The medium fine grid is generated using the same method based on the medium grid. The fine and Extra fine grids are mainly refined along the flow and span directions. Adding more grid points in the thickness direction can lead to computational instability. The cell numbers are approximately 149 million, 240 million, 333 million, and 493 million for the medium, medium fine, fine, and extra fine grids, respectively. Fig. 16 presents the surface grid of the medium grid, in which the $x$ and $y$ directions are the streamwise direction and the spanwise direction. The first grid layer of the five grid sets has $\Delta y^+$ values of approximately 1.0, 0.6, 0.4, 0.2, and 0.2, respectively.

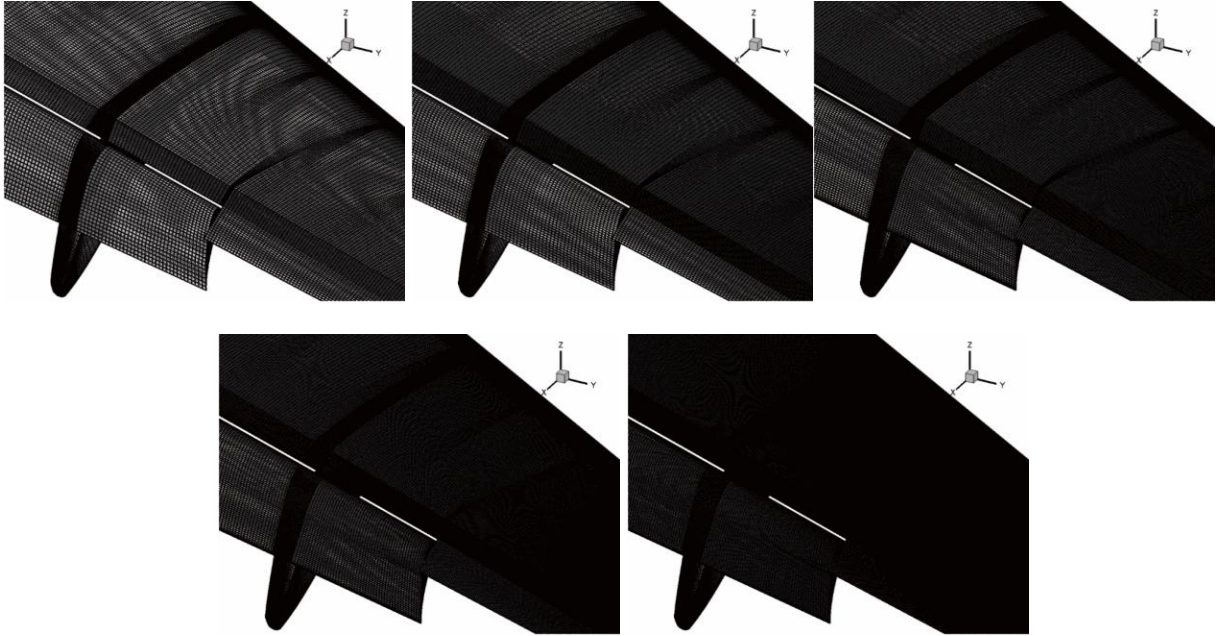

**Fig. 15 Surface grids near the side edges of flaps with different grids**

**Table 2 Grid parameters with different densities**

| Grid name | Cell number (million) | $y^+$ | $C_{L,max}$ |
|---|---|---|---|
| **Coarse** | 78 | 1.0 | 2.26 |
| **Medium** | 149 | 0.6 | 2.39 |
| **Medium Fine** | 240 | 0.4 | 2.44 |
| **Fine** | 333 | 0.2 | 2.45 |
| **Extra Fine** | 493 | 0.2 | 2.43 |

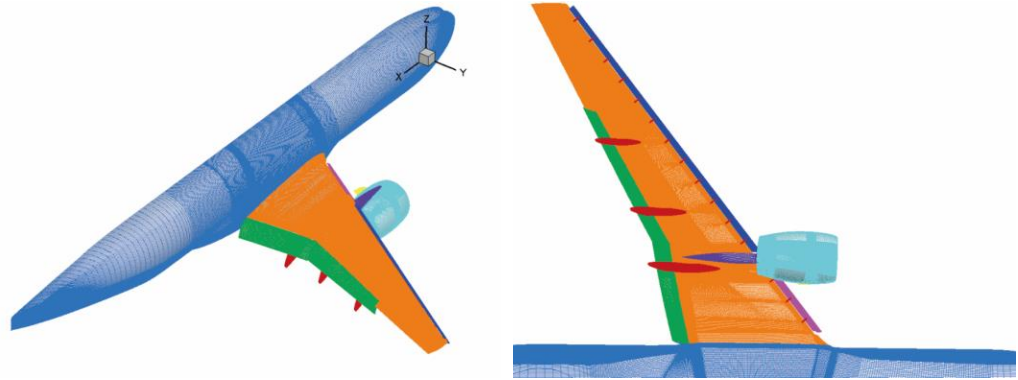

(b) Top view  (b) Bottom view

**Fig. 16 Surface grid of the CRM-HL medium grid**

Fig. 17 presents the results of the grid refinement study using the SPF $k - \overline{v^2} - \omega$ model. The coarse grid underpredicts the maximum lift coefficient and the stall angle of attack. The results from medium fine grids show a slight improvement over those from medium grids. The medium fine, fine and extra fine grids produce nearly identical results at high angle of attack, and they agree well with the experimental data. However, the instability during the convergence process is more noticeable with the fine and extra fine grids beyond maximum lift, as shown in Fig. 17(d). The final time step has to be reduced to 0.5 to ensure computational stability. Regarding the drag coefficient, the coarse grid predicts results similar to the medium and medium fine grids when the lift coefficient is below approximately 1.2. However, when the lift coefficient exceeds 1.2, the coarse grid overpredicts the drag coefficient. The medium and medium fine grids provide better drag predictions before stall, and they slightly overpredict the drag coefficient near stall. The drag obtained using the fine and extra fine grids is higher after stall compared to the results from the medium fine grid. The pitching moments predicted by the coarse grid deviate from the experimental data across the entire range of angle of attack values. The results obtained using the medium, medium fine and fine grids are almost the same before stall, and their tendencies match well with the experimental data. The results from the Extra fine grid show a slight deviation compared to these results. In summary, the SPF $k - \overline{v^2} - \omega$ model shows satisfactory grid convergence in predicting the aerodynamic performance of the CRM-HL configuration. The medium fine grid is used for the subsequent numerical simulations to accurately assess the differences in stall behavior predicted by different models.

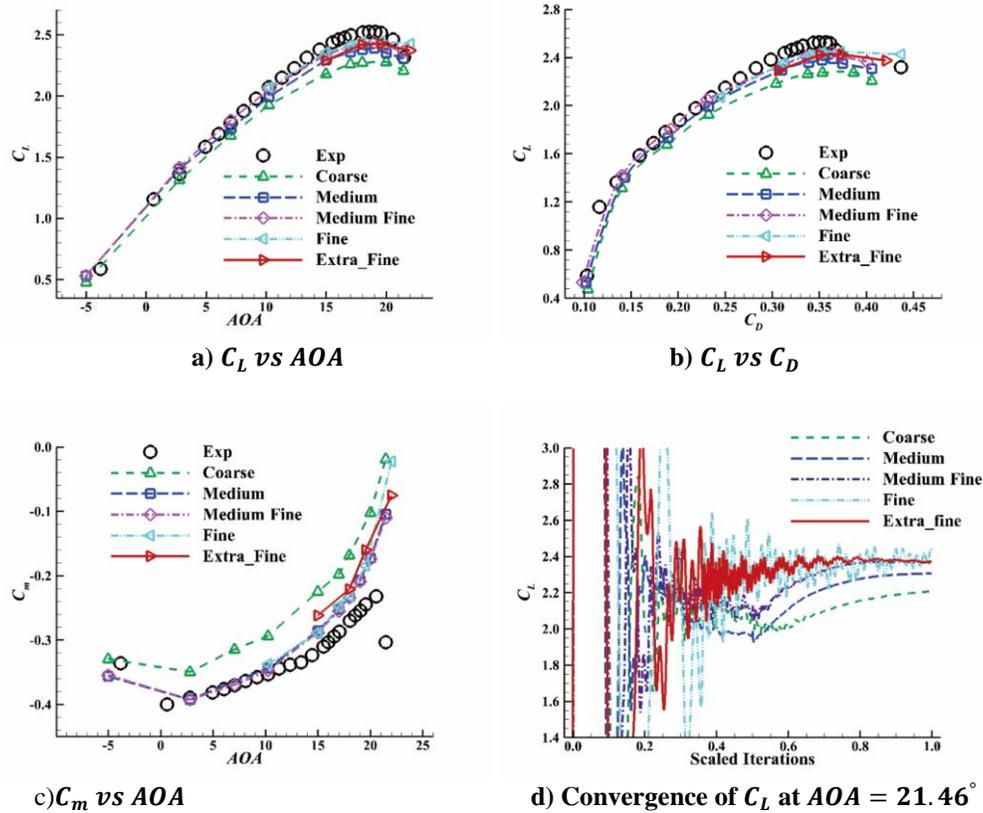

a) $C_L$ vs AOA

b) $C_L$ vs $C_D$

c) $C_m$ vs AOA

d) Convergence of $C_L$ at $AOA = 21.46°$

**Fig. 17 Grid refinement study for the CRM-HL configuration using the SPF $k - \overline{v^2} - \omega$ model**

Fig. 18 displays the lift coefficient in relation to $h$, indicating the spacing between mesh points. The mesh spacing gets finer as h approaches 0. SST predicts stall at $AOA = 18°$, and the lift coefficient approaches near 2.2. The maximum lift angle of attack predicted by the SA and original $k - \overline{v^2} - \omega$ models is around 18 degrees. The predicted lift coefficients head toward 2.3 and 2.35, respectively. The original $k - \overline{v^2} - \omega$ model couldn't reach a convergent solution with an extra fine mesh. but the SPF model shows improved convergence. Lift coefficient predicted by the SPF model head toward near 2.43. Convergence of steady state can be obtained using the medium fine grid with SA, original $k - \overline{v^2} - \omega$, and SPF models. The SST model cannot obtain a steady-state convergent solution at this angle of attack, as shown in Fig. 19 (a). The unsteady state is more evident with the Extra fine grid, where the lift fluctuates within a specific range after achieving convergence.

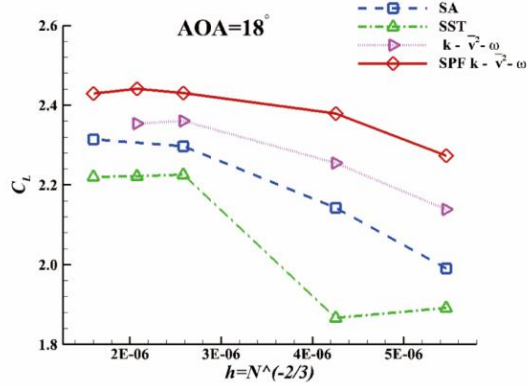

**Fig. 18 Mesh convergence of CRM-HL lift coefficient using different turbulence models at $AOA = 18°$**

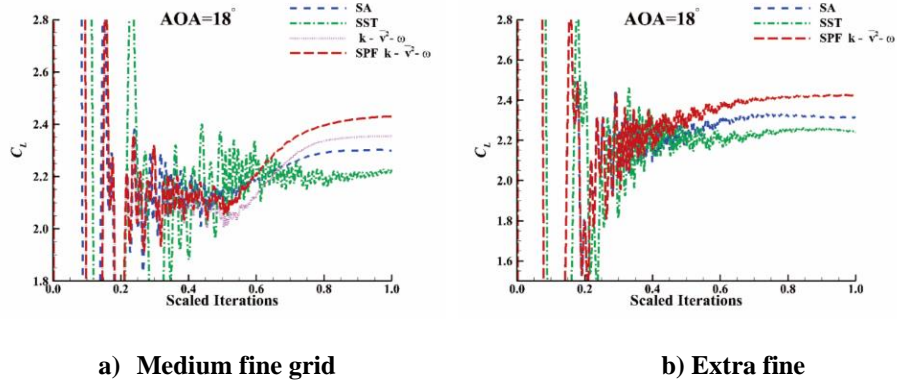

   a) **Medium fine grid**          b) **Extra fine**

**Fig. 19 Scaled iterative convergence of $C_L$ for different turbulence models**

*4. Results with chine*

 Using the medium fine grid for computing leads to better convergence in solutions. The simulation in this section is carried out on the basis of medium fine grid. The aerodynamic coefficients predicted by the SA, SST, original $k - \overline{v^2} - \omega$, SPF $k - \overline{v^2} - \omega$ and SPF-RC models are shown in Fig. 20. The maximum lift coefficient of CRM-HL is underpredicted by the SST and SA models. The original $k - \overline{v^2} - \omega$ model outperforms the two models in predicting the maximum lift coefficient. However, the stall angle predicted by the original $k - \overline{v^2} - \omega$ model occurs 1.0 degrees earlier than the experimental data. The SPF $k - \overline{v^2} - \omega$ model provides a more accurate prediction of the maximum lift coefficient compared to the original model, and it predicts the same stall angle as the experimental data. The SPF-RC model predicts a slightly higher maximum lift coefficient than the SPF model, and it also shows better agreement with the experimental data. The relative errors of the maximum lift coefficient predicted by the SA, SST, original $k - \overline{v^2} - \omega$, SPF $k - \overline{v^2} - \omega$ and SPF-RC models are 11.4%, 8.7%, 6.4%, 3.3%, and 2.8%, respectively. The five models predict nearly the same drag coefficient values before stall. Compared with the experimental data near stall, the SPF $k-\overline{v^2}-\omega$ and SPF-RC models more accurately predict the drag coefficient. In terms of the pitching moment, the results predicted by the SA and SST models deviate from the experimental data. The original $k-\overline{v^2}-\omega$ model and the SPF models predict results that are very close to the experimental data when the angle of attack values are less than 10 degrees. The SPF model exhibits better agreement with experiments at high angle of attack values, with only a minor difference compared to the predictions of the SPF-RC model.

The SA turbulence model for RANS is used by most participants in the Fourth AIAA High Lift Prediction Workshop. The outcomes of the two "best practices" studies are detailed in references [41] and [42]. They obtain a more perfect maximum lift coefficient compared to this study, as shown in Fig. 20(a). The SPF and SPF-RC models predicted the pitching moment coefficient more accurately. However, they couldn't replicate the "Pitch break" effect observed in experiments and WMLES results beyond $C_{L,max}$ [39, 43].

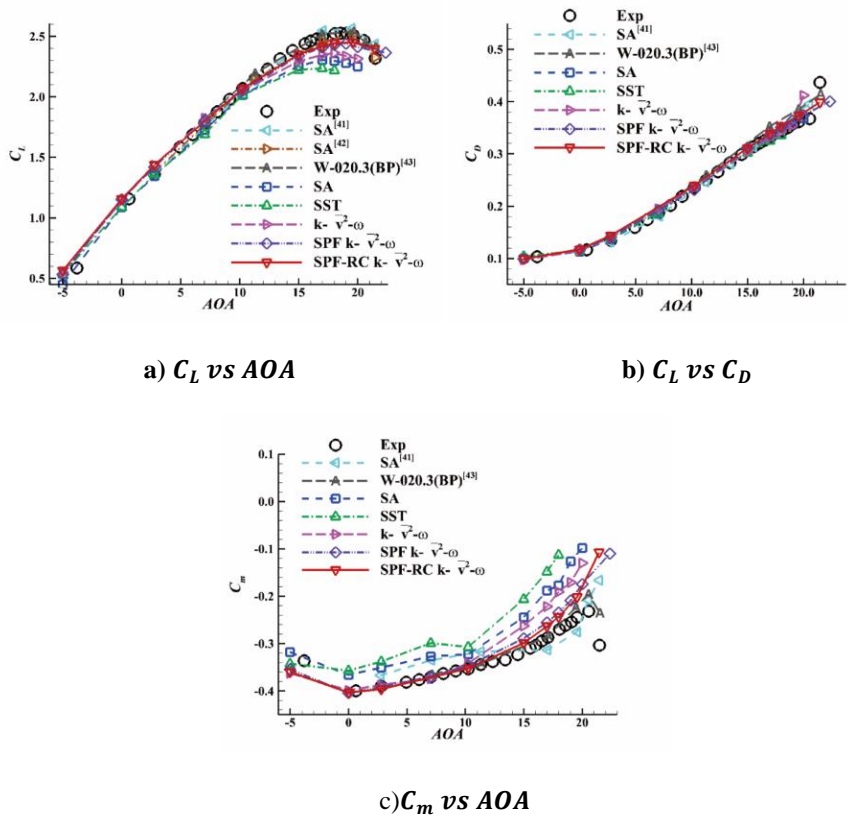

a) $C_L$ vs AOA

b) $C_L$ vs $C_D$

c) $C_m$ vs AOA

**Fig. 20 Aerodynamic coefficients predicted by different models of the CRM-HL configuration**

The corrected lift coefficient from wind tunnel tests at $AOA = 7.05$ is 1.78. The predicted lift coefficients for SA, SST, $k - \overline{v^2} - \omega$, and SPF models are 1.73, 1.69, 1.81, and 1.80, respectively. Both the SA and SST models underestimated the lift coefficient and predicted a more noticeable outboard flap separation. The peak suction on the main and flap elements at section D-D, E-E, and F-F are significantly unpredicted. This phenomenon is also observed in the results of R-025.3 [40]. The original $k - \overline{v^2} - \omega$ model and SPF model produced nearly identical results, with peak suction on the main and flap elements aligning closely with experimental data.

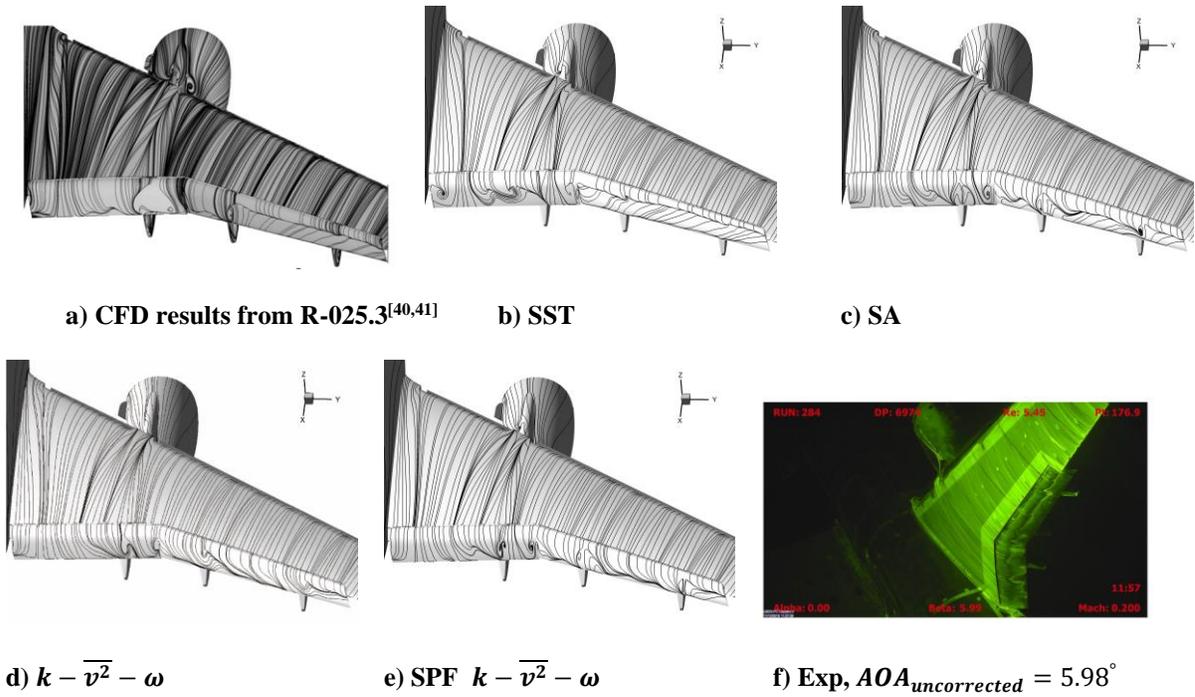

a) CFD results from R-025.3[40,41]    b) SST    c) SA

d) $k - \overline{v^2} - \omega$    e) SPF $k - \overline{v^2} - \omega$    f) Exp, $AOA_{uncorrected} = 5.98°$

**Fig. 21 Surface streamlines predicted by different models and oil flow visualization obtained from the wind tunnel experiment of CRM-HL, $AOA = 7.05°$**

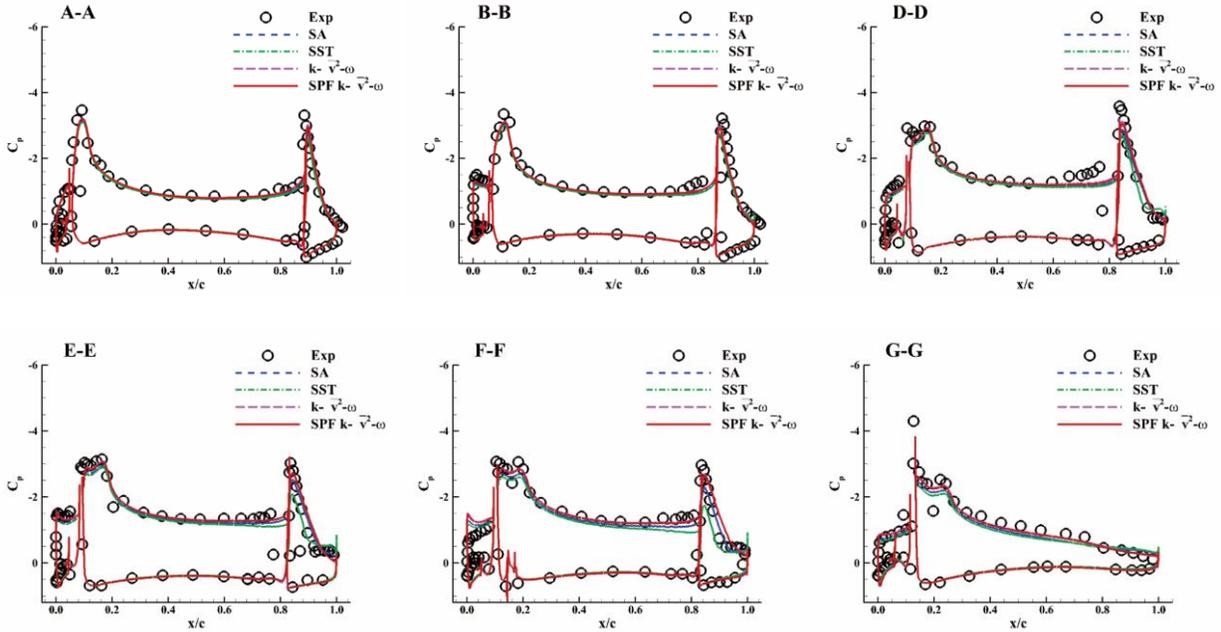

**Fig. 22 Pressure distributions of the CRM-HL configuration at different span locations obtained by different models, AOA = $7.05°$**

The results under conditions close to stall are analyzed here. Sections at different span locations and the chine vortex illustrated by streamline are shown in Fig. 23. The SPF $k - \overline{v^2} - \omega$ model captures the direction change of the chine vortex, which curves toward the wing's root as it traverses the upper wing surface. This behavior is in line with the results from wall-modeled large eddy simulations (WMLES) by Wang et al.[45].

Fig. 24 displays pressure distributions of the CRM-HL configuration at different span locations at AOA=17 degrees. The SA and SST models underpredicted suction peaks, especially the suction peaks of the flaps along Sections B-B and F-F. The pressure distributions on the trailing edge of the main wing at these two sections deviated more from the experimental data. The original $k - \overline{v^2} - \omega$ model performs better than the SA and SST models and predicts a more accurate suction peak. The SPF $k - \overline{v^2} - \omega$ model further enhances the pressure distribution. The SPF-RC model slightly improves suction peaks at Sections D-D and E–E than the SPF model.

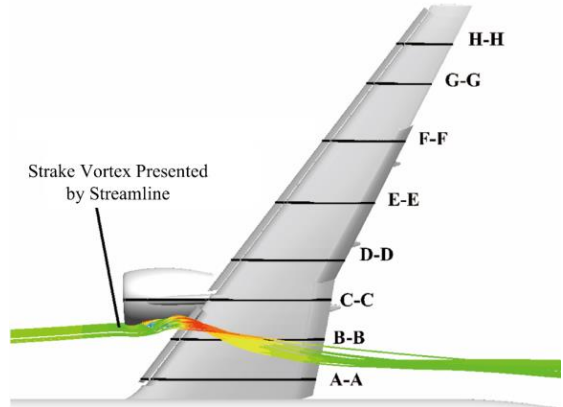

**Fig. 23 Section location and chine vortex presented by streamline predicted by the SPF $k - \overline{v^2} - \omega$ model**

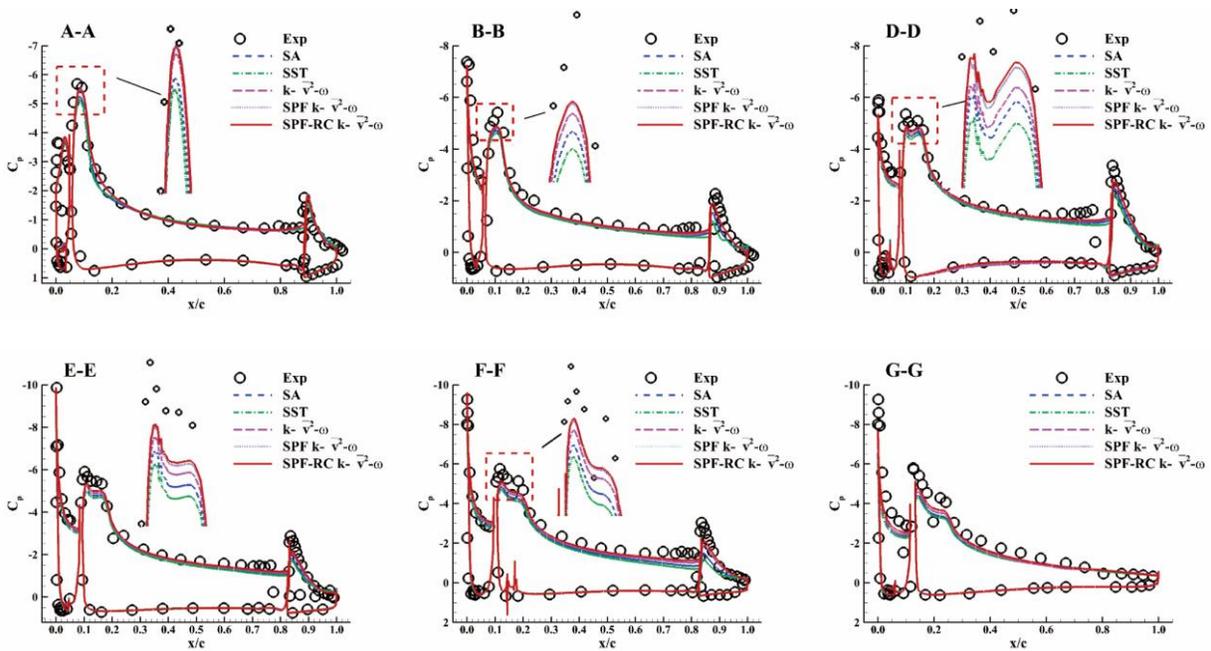

**Fig. 24 Pressure distributions of the CRM-HL configuration at different span locations obtained by different models, AOA = $17°$**

Fig. 25 presents the nondimensional velocities $U/U_{inf}$ at Section B-B obtained by different models. Section B-B is particularly affected by the chine vortex, which helps reduce the low-speed area caused by sheltering from the pylon/nacelle. A free shear layer forms between the primary flow area and the low-speed region sheltered by the pylon/nacelle due to strong velocity gradients. The SA and SST models predict larger main wing wakes than the original $k - \overline{v^2} - \omega$ model and the SPF models. The main wing wakes above the flap tend to suppress the flap's suction peak [47]. This explains why the SA and SST models predict a lower flap suction peak in this section. The

**26/40**

SPF model performs slightly better than the original $k - \overline{v^2} - \omega$ model, forecasting a smaller low-speed region above the trailing edge of the main wing. The SPF $k - \overline{v^2} - \omega$ model predicts a higher flow velocity in the core of the chine vortex than the SA, SST, and original $k - \overline{v^2} - \omega$ models. The vortex core predicted by the SPF-RC model shifts slightly downstream when compared to the SPF model results presented in this section.

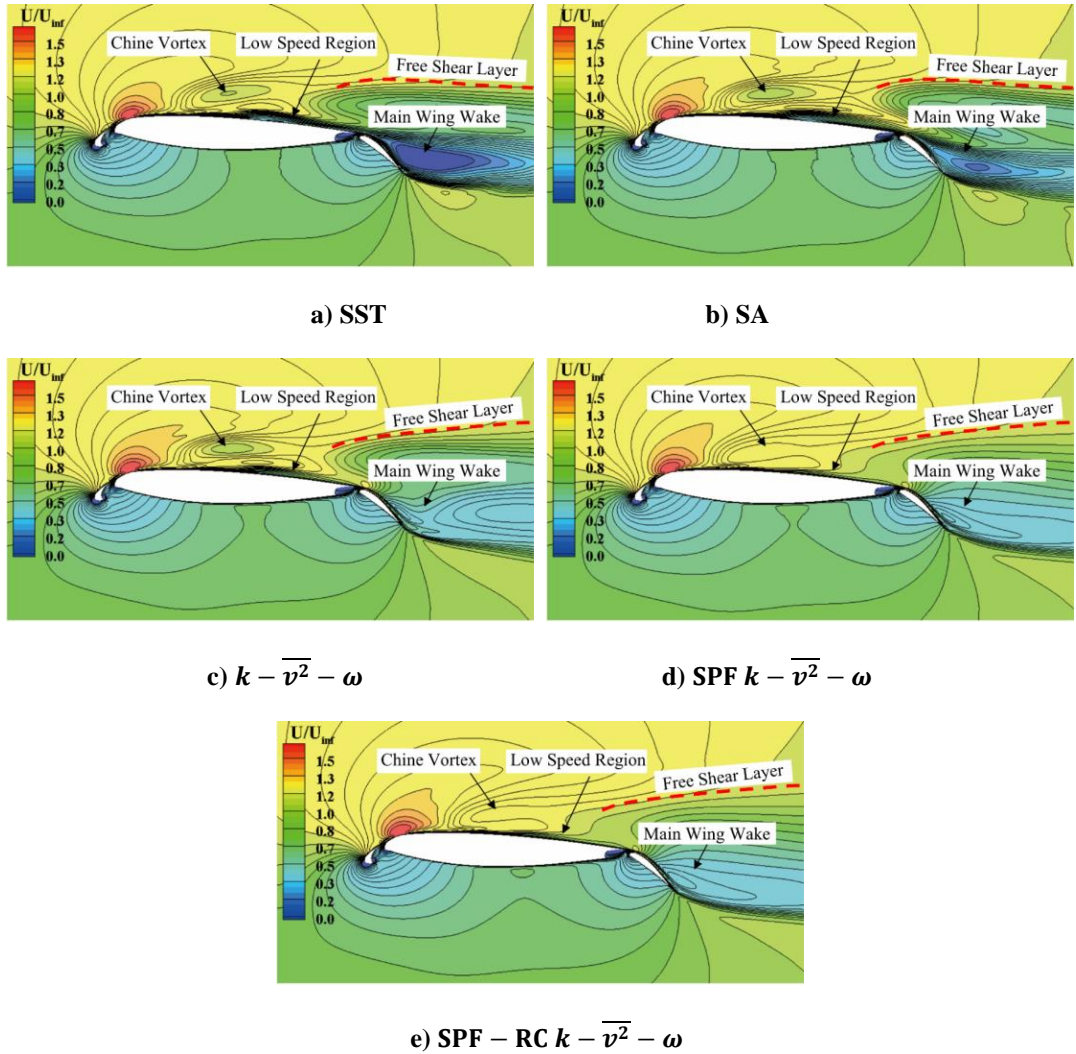

a) SST

b) SA

c) $k - \overline{v^2} - \omega$

d) SPF $k - \overline{v^2} - \omega$

e) SPF $-$ RC $k - \overline{v^2} - \omega$

**Fig. 25 Comparison of the $U/U_{inf}$ contours at Section B-B obtained by different models, AOA $= 17°$**

Fig. 26 presents a comparison of the $P_{\overline{v^2}}/\varepsilon$ contours at Section B-B obtained using the $k - \overline{v^2} - \omega$ and SPF $k - \overline{v^2} - \omega$ models at AOA $= 17°$. The production-to-dissipation ratio $P_{\overline{v^2}}/\varepsilon$ is obtained using the original $k - \overline{v^2} - \omega$ and SPF models according to Eq. (5). The value of $P_{\overline{v^2}}/\varepsilon$ predicted by the original $k - \overline{v^2} - \omega$ model is notably higher than 1.5, reaching approximately 2.5 in the free shear layer at specific locations. The modification for the separated shear layer is activated when $P_{\overline{v^2}}/\varepsilon$ exceeds 2.5. The SPF model significantly increases $P_{\overline{v^2}}/\varepsilon$ in the free shear layer, as depicted in Fig. 18(b). The turbulence intensity on the lower surface of the flap is quite low, resulting in a relatively large $P_{\overline{v^2}}/\varepsilon$ value.

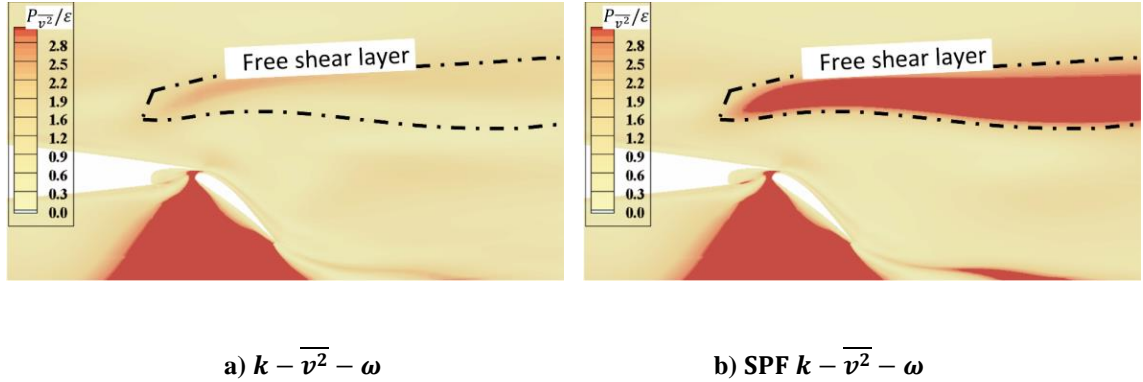

a) $k - \overline{v^2} - \omega$  b) SPF $k - \overline{v^2} - \omega$

**Fig. 26 Comparison of the $P_{\overline{v^2}}/\varepsilon$ contours at Section B-B obtained by the $k - \overline{v^2} - \omega$ and SPF $k - \overline{v^2} - \omega$ models, AOA = 17°**

The evolutions of the chine, outboard and nacelle/pylon vortices predicted by different models at AOA = 17° are shown in Fig. 27. The values of the *x*-component of vorticity are high near the chine, outboard, and nacelle/pylon vortex positions. All models predict a strong chine vortex that extends beyond the trailing edge. The SA, SST, and original $k - \overline{v^2} - \omega$ models predict similar intensities for the chine and nacelle/pylon vortices. However, the SPF model predicts a weaker chine vortex with a broader influence, as shown in Figure 19(d). This difference is due to an increase in the production-to-dissipation ratio in the free shear layer. The enhanced turbulence accelerates the exchange of energy between the mainstream region and the low-speed region, resulting in weakened vortex intensity and faster diffusion. In the case of the $k - \overline{v^2} - \omega$-RC and SPF-RC models, the chine vortex intensity values are slightly increased, while other vortex intensities remain largely unchanged when compared to the results predicted by the SPF model.

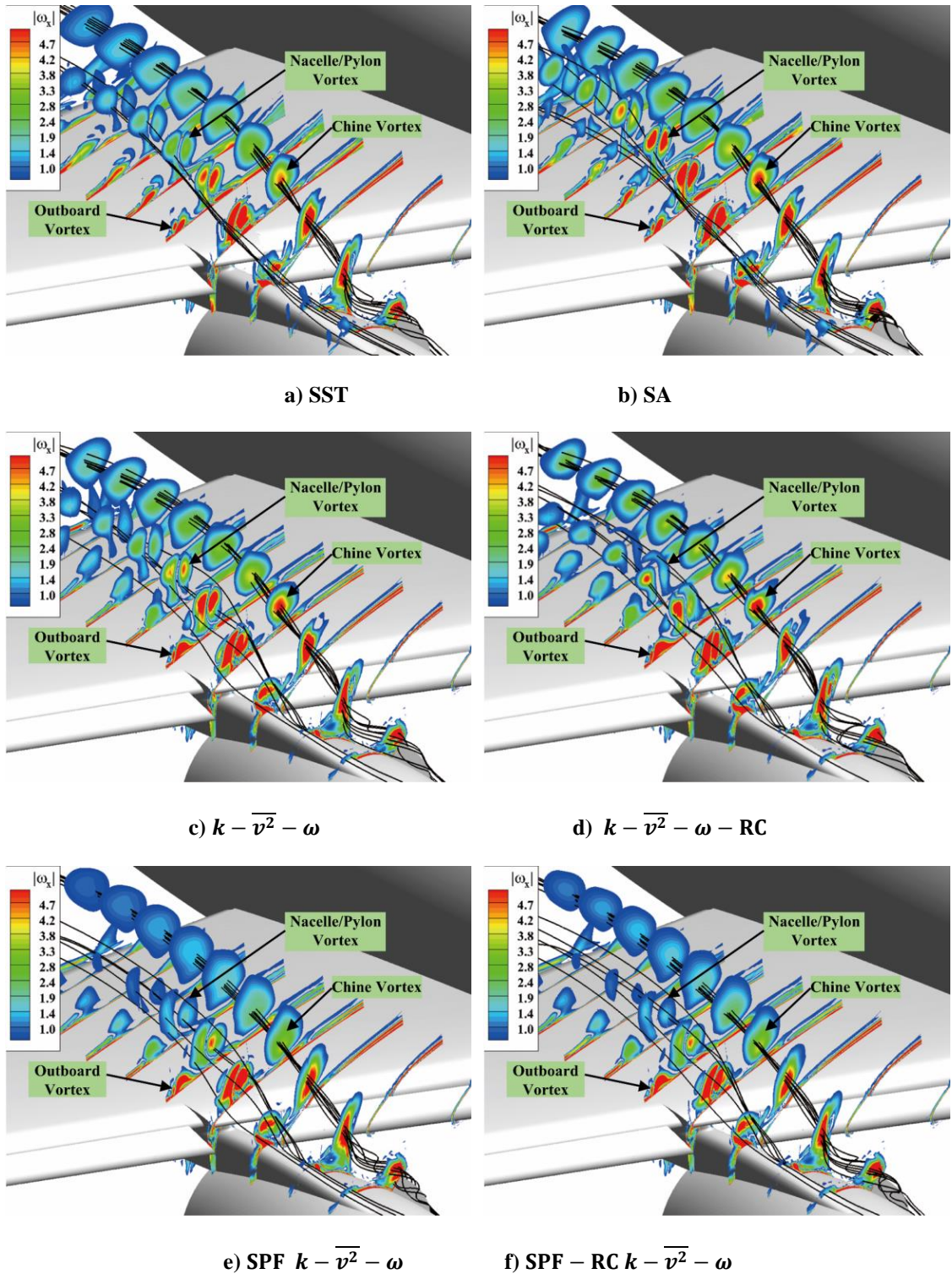

**Fig. 27 Chine, outboard, and nacelle/pylon vortices predicted by different models, AOA = 17°**

Surface friction coefficient predicted by different models and oil flow visualization obtained from the wind tunnel experiment are shown in Fig. 28. Vortices generated by the slat brackets lead to four noticeable separation regions on the outboard section of the wing. The SA and SST models predicted an earlier stall, with results shown at $AOA = 18°$. All these models predict less separation as the grid is refined from coarse to medium fine. The predicted separation patterns of these models show no significant differences when the grid changes from medium fine to extra fine. The SPF model demonstrates strong grid convergence, with the medium fine, fine, and Extra fine grids yielding similar surface flow patterns. In contrast, the original $k - \overline{v^2} - \omega$ model failed to reach convergence with fine and Extra fine meshes, indicating that SPF corrections can improve the model's convergence. The SA and SST models predict relatively large outboard separation regions. This phenomenon is also observed in the results of A-025.1 [40]. The original $k - \overline{v^2} - \omega$ model predicts a somewhat large separation on the wing outboard using the medium fine grid. The SPF $k - \overline{v^2} - \omega$ model offers separation patterns that closely align with the experimental results. Adding rotation correction to the SPF model has little effect on the separation characteristics of the wing outboard, as shown in Fig. 28(d).

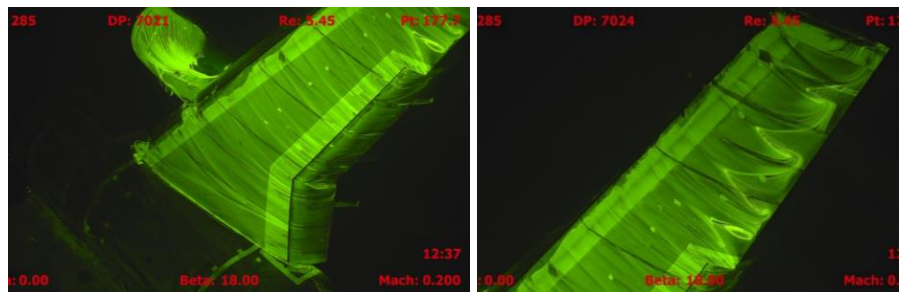

(a) **Exp,** $AOA_{uncorrected} = 17.98°$ ($AOA_{corrected} = 19.57°$)

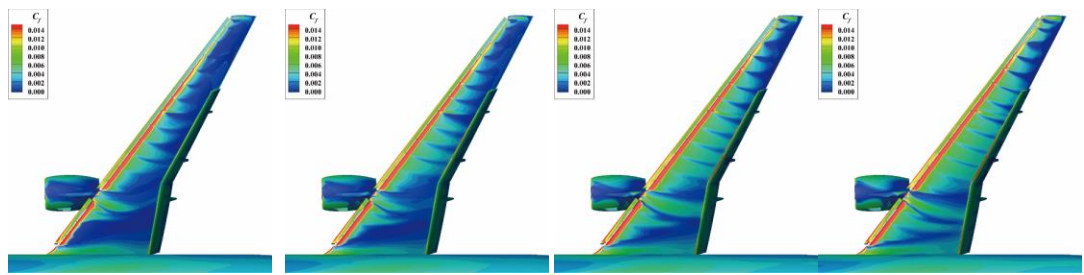

1) Coarse    2) Medium    3) Fine    4) Extra fine plus

(b) SA, $AOA = 18°$

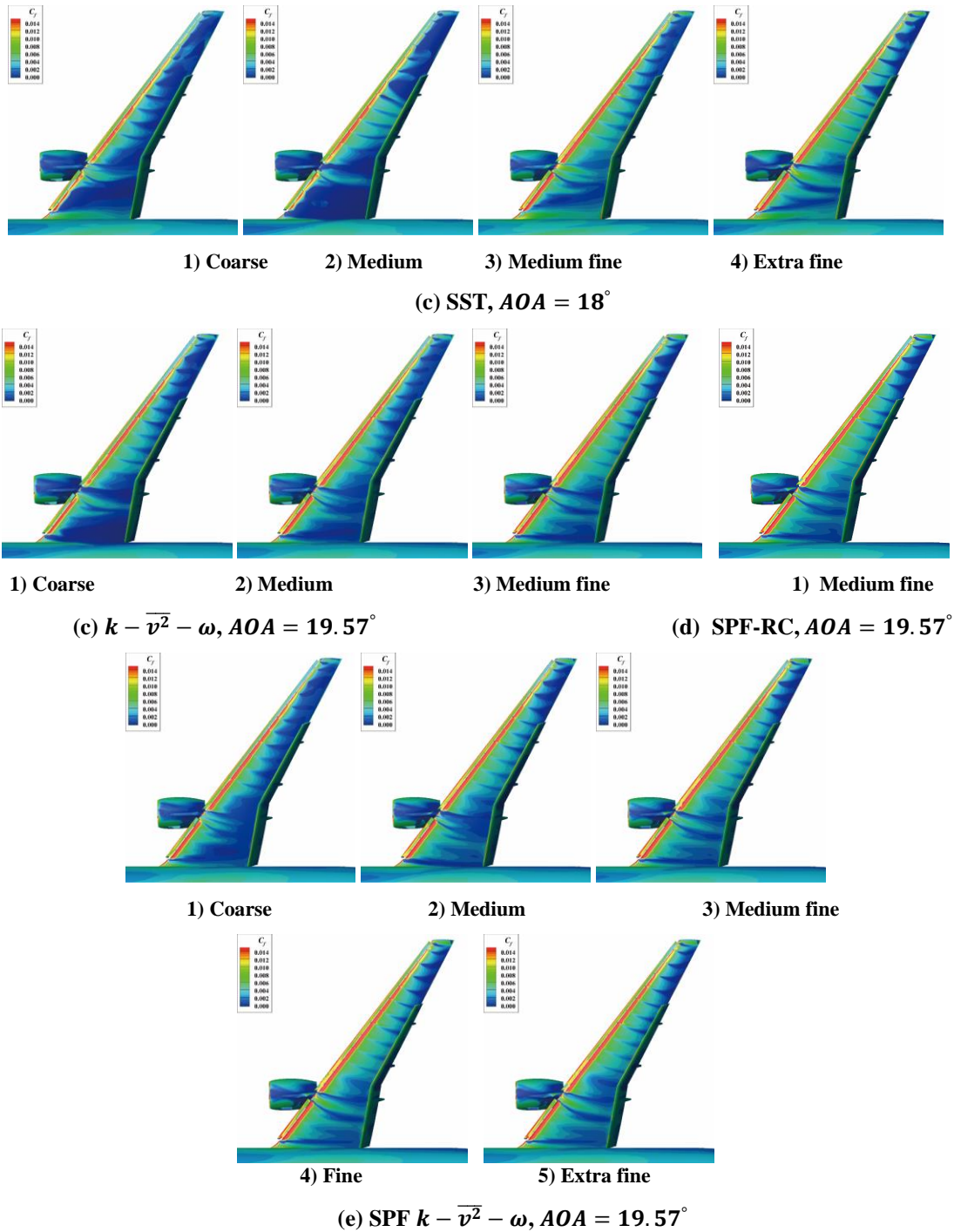

**Fig. 28 Surface streamlines and pressure coefficient contours predicted by different models and oil flow visualization obtained from the wind tunnel experiment for the wing outboard of CRM-HL**

The SPF model predicts smaller separation bubbles at section I-I position, as shown in Fig. 29. The separation on the wing outboard caused by the vortices formed by the slat brackets at high angles of attack. The low-speed region inside the separation bubble forms a distinct shear layer upon encountering the high-speed airflow from the main flow region, as shown in Fig. 30. The modification for the separated shear layer is activated when $P_{\overline{v^2}}/\varepsilon$ exceeds 2.5. The SPF model significantly increases $P_{\overline{v^2}}/\varepsilon$ in the separated shear layer, as depicted in Fig. 30 (b). The enhanced turbulence accelerates the exchange of energy between the mainstream region and the separated region, resulting in weakened vortex intensity and faster diffusion.

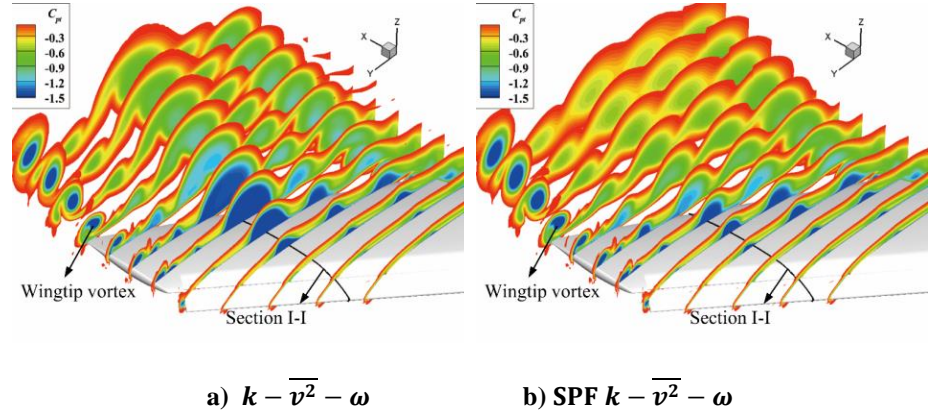

a) $k - \overline{v^2} - \omega$      b) SPF $k - \overline{v^2} - \omega$

**Fig. 29 Comparations of the total pressure coefficient contours of the CRM-HL configuration, AOA = 18°**

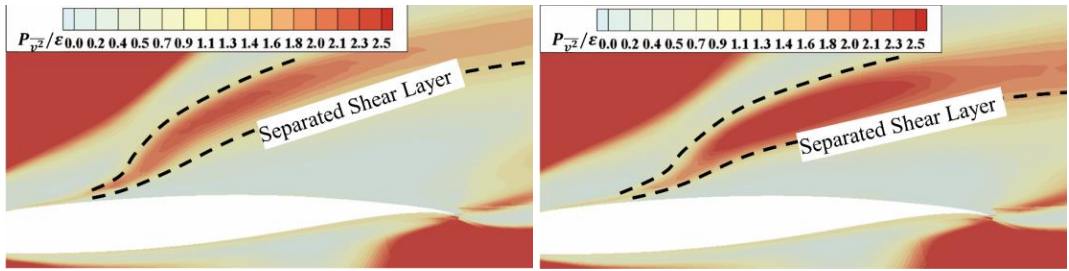

**Fig. 30 Comparison of the $P_{\overline{v^2}}/\varepsilon$ contours at Section I-I obtained by the $k - \overline{v^2} - \omega$ and SPF $k - \overline{v^2} - \omega$ models, AOA = 18°**

*5. Results without chine*

In this section, the impact of the nacelle chine is numerically investigated. The results provided by the SPF-RC model demonstrate better alignment with the experimental data, as discussed above. Therefore, all calculations in this section are based on this model. Fig. 31 illustrates the aerodynamic coefficients predicted by the SPF-RC $k - \overline{v^2} - \omega$ model for the CRM-HL configuration with and without the nacelle chine. The linear segments of the lift curves for configurations with and without the nacelle chine are nearly identical. However, notable differences emerge near the stall region. The maximum lift coefficients for configurations with and without the nacelle chine measure 2.44 and 2.39, respectively. The presence of the chine reduces the drag coefficient of the CRM-HL configuration at high angle of attack values, and variations in the pitching moment coefficient are primarily observed at high angle of attack values.

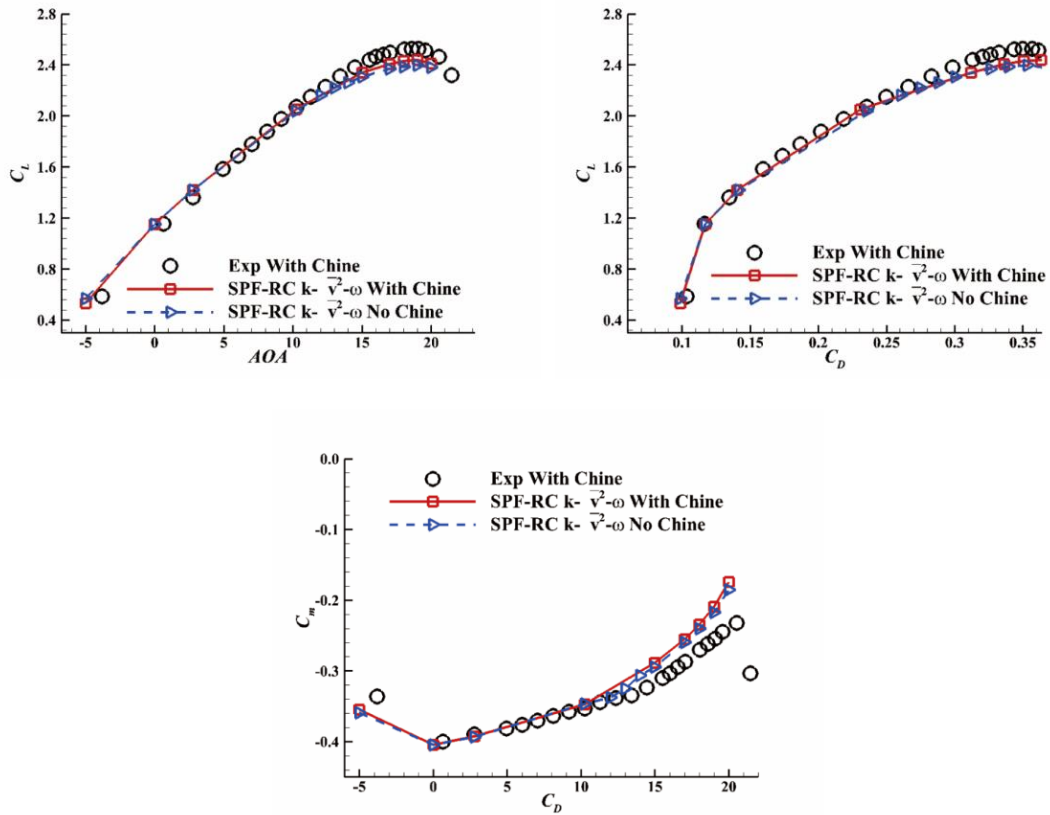

**Fig. 31 Comparations of the aerodynamic coefficients of the CRM-HL configuration with and without the nacelle chine**

Fig. 32 presents a comparison of the total pressure coefficient contour images ($C_{pt} = 2(p_t - p_{t\infty})/(\rho V_\infty^2)$) for the CRM-HL configuration with and without the nacelle chine. Negative values in the total pressure indicate a deficiency primarily caused by the vortex system. Vortices originating from the cutout region near the pylon/nacelle for the inboard and outboard slats are referred to as the inboard vortex and outboard vortex, respectively. These two vortices combine to create the nacelle/pylon vortex system. The chine vortex plays a significant role in mitigating the impact of the inboard vortex, effectively eliminating the low-energy region downstream.

Surface flow visualization images with and without a nacelle chine are shown in Fig. 33. Fig. 33(a-b) shows the skin friction coefficient predicted by the SPF-RC model of the CRM-HL configuration with and without the nacelle chine. A negative skin friction coefficient indicates separation of the wing surface. The separation of the wing inboard gradually increases with increasing angle of attack, which causes the high-lift configuration to stall. The flow separation is dominated by the nacelle/pylon vortex system when the chine is not present. In contrast, flow separation is mainly dominated by the wing/body separation bubble when the chine is present. The chine vortex can weaken the adverse effect and reduce the separation region induced by the nacelle/pylon vortex system, as demonstrated by results obtained by the SPF-RC model. This phenomenon is consistent with Koklu et al. [48].

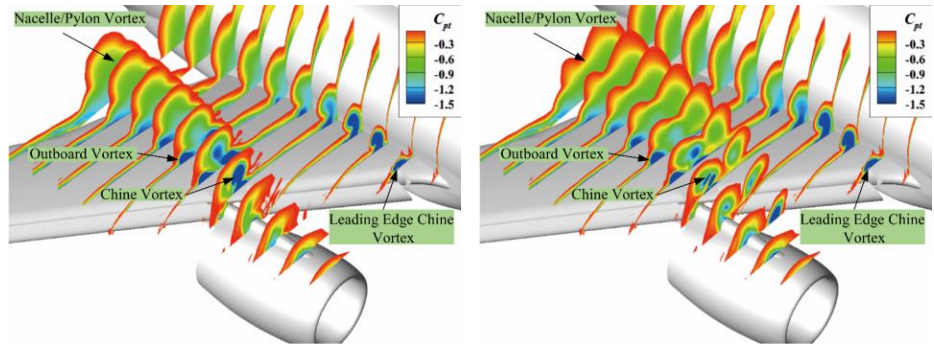

(a) SPF-RC $k - \overline{v^2} - \omega$, without chine  (b) SPF-RC $k - \overline{v^2} - \omega$, with chine

**Fig. 32 Comparations of the total pressure coefficient contours of the CRM-HL configuration with and without the nacelle chine, $AOA = 19.57°$**

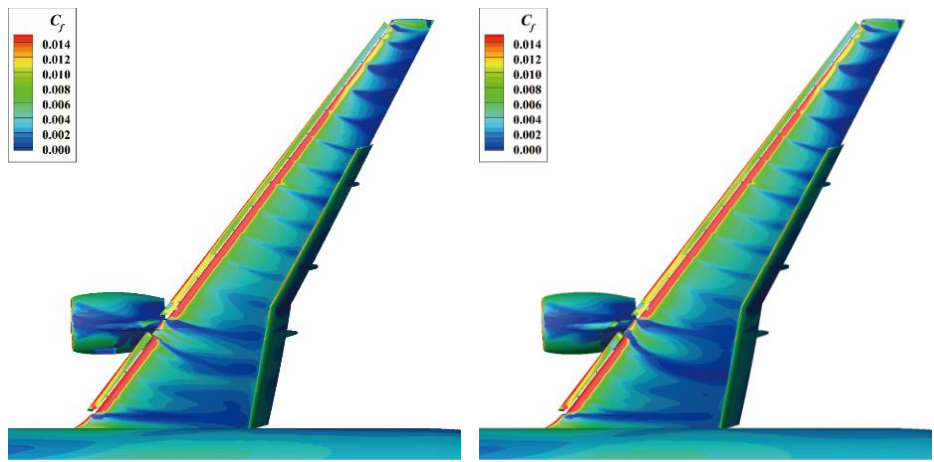

(a) SPF $k - \overline{v^2} - \omega$, With Chine, $AOA = 19.57°$  (b) SPF $k - \overline{v^2} - \omega$, No Chine, $AOA = 19.57°$

**Fig. 33 Surface flow visualization for the CRM-HL configuration with and without the nacelle chine**

Section B-B experiences the most significant impact from the nacelle/pylon vortex system. Comparisons of the $U/U_{inf}$ contours at Section B-B obtained using the SPF-RC model of the CRM-HL configuration with and without the nacelle chine are presented in Fig. 34. The nacelle/pylon creates a substantial low-speed area above the trailing edge of the main wing due to its sheltering effect in the absence of the chine. This low-speed region combines with the main wing wake, resulting in an extensive low-speed region. However, the chine vortex counteracts this effect. It reduces both the low-speed region above the trailing edge of the main wing and the extent of the main wing wake.

This reduction in the low-speed area enhances the overall circulation, which is the reason for the nacelle chine improving the stall performance.

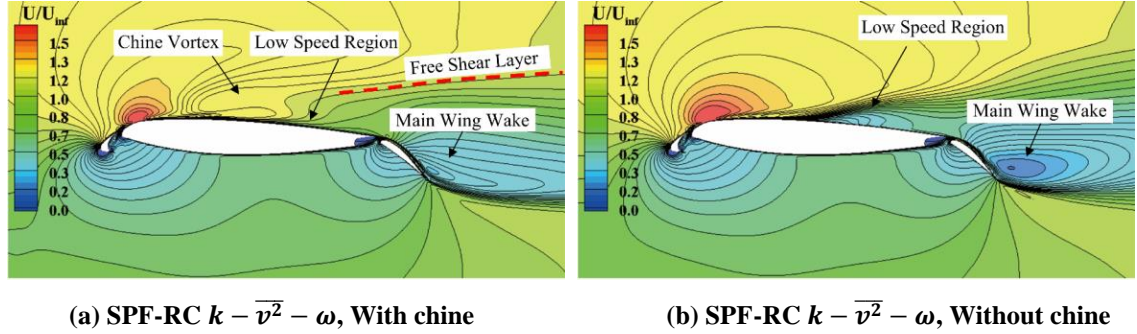

(a) SPF-RC $k - \overline{v^2} - \omega$, With chine  (b) SPF-RC $k - \overline{v^2} - \omega$, Without chine

Fig. 34 Comparison of the $U/U_{inf}$ contours at Section B-B obtained by the SPF-RC $k - \overline{v^2} - \omega$ of the CRM-HL configuration with and without the nacelle chine, $\text{AOA} = 19.57°$

## IV. Conclusion

The study presented in this paper conducted aerodynamic predictions for the high-lift common research model used in the fourth AIAA High-Lift Prediction Workshop. Several Reynolds-averaged Navier–Stokes (RANS) models were employed to study the stall behavior of the configuration, including the SA, SST, original $k - \overline{v^2} - \omega$, SPF $k - \overline{v^2} - \omega$, and SPF-RC models. The work can be summarized as follows.

(1) A vortex generator is numerically studied to validate the impact of separated shear layer fixed and rotation corrections on the model's ability to predict the vortex strength of a vortex generator. The original $k - \overline{v^2} - \omega$ model with separated shear layer correction does not affect the prediction of vortex strength. The SPF model, with the addition of rotation correction, can improve the model's prediction of vortex strength.

(2) The multielement airfoil 30P30N is chosen as the second case to test the influence of these two corrections on the stall behavior prediction of the multielement airfoil. The SST and original $k - \overline{v^2} - \omega$ models underpredict the maximum lift coefficient and predict more pronounced separation bubbles near the stall angle of attack. The SA and SPF $k - \overline{v^2} - \omega$ models predict slight flow separation, resulting in better agreement with experimental observations. Rotation correction of the SPF model has almost no effect on the prediction of stall performance.

(3) The complex three-dimensional model tested in study presented in this paper is the CRM-HL configuration with a nacelle chine. The SA and SST models fail to predict the stall behavior of this configuration and predict large separation bubbles on the wing outboard. The SPF $k - \overline{v^2} - \omega$ more accurately predicts the stall behavior than the original model. The flow pattern on the wing outboard predicted by the SPF $k - \overline{v^2} - \omega$ model is consistent with the experimental data. The SPF-RC model predicts a stronger vortex intensity of the nacelle chine, and the predicted maximum lift coefficient is further improved. The relative errors in predicting the maximum lift coefficient of the SPF-RC $k - \overline{v^2} - \omega$ model are approximately 2.8% of the experimental data. Compared to the RANS results of the HLPW4 "best practice", the SPF-RC model predicts more accurate pitching moment coefficient, forecasts less outboard flap separation at $AOA = 7.05°$, and predicts more reasonable outboard wing separation near stall.

(4) The effect of the nacelle chine on the stall behavior of CRM-HL is studied. A large low-speed region is formed above the trailing edge of the main wing for the no-chine configuration. A chine vortex can reduce the effect of the nacelle/pylon vortex and make the boundary layer of the main wing healthier. The configuration with the nacelle chine can increase the maximum lift coefficient by approximately 2% when compared to the configuration without the nacelle chine.

## Acknowledgments

This work was supported by the National Natural Science Foundation of China (grant nos. 12372288, 12388101, U23A2069, and 92152301).